\documentclass[preprint2]{aastex}

\newcommand{\kms}{\mbox{km/s}}

\newcommand{\kpc}{\mbox{kpc}}
\newcommand{\pc}{\mbox{pc}}

\newcommand{\Msun}{\mbox{M$_\odot$}}
\newcommand{\msun}{\mbox{M$_\odot$}}

\begin{document}

\title{The rotation curves of dwarf galaxies: a problem for Cold Dark Matter?} 
\author{George Rhee} 
\affil{Physics Department, University of Nevada, \\ 
Box 454002, Las Vegas, NV 89154} 
\and
\author{Octavio Valenzuela} 
\affil{Astronomy Department, University of Washington \\
Box 351580, Seattle, WA 98195}
\author{Anatoly Klypin, Jon Holtzman and Bhasker Moorthy} 
\affil{Astronomy Department, New Mexico State University, \\
PO Box 30001, Las Cruces NM 88003}

\keywords{
galaxies: halos---galaxies: local group---galaxies: dwarf---galaxies: 
kinematics and dynamics---cosmology: dark matter---}

\begin{abstract}

We address the issue of accuracy in recovering density
profiles from observations of rotation curves of galaxies. We
``observe'' and analyze our models in much the same way as observers
do the real galaxies.  Our models include stellar disks, disks with
bars, and small bulges. We find that the tilted ring model analysis
produces an underestimate of the central rotational velocity. In some
cases the galaxy halo density profile seems to have a flat core, while
in reality it does not.  We identify three effects, which explain the
systematic biases: (1) inclination (2), small bulge, and (3) bar.
Inclination effects are due to finite thickness of disk, bar, or
bulge.  Admixture of a non-rotating bulge component reduces the
rotation velocity. A small (200-500~pc) bulge may be overlooked
leading to systematic bias even on relatively large $\sim 1~\kpc$
distances.  In the case of a disk with a bar, the underestimate of the
circular velocity is larger due to a combination of non-circular
motions and random velocities.  The effect of the bar depends on the
angle that the bar makes with the line of sight. Signatures of bars
can be difficult to detect in the surface brightness profiles of the
model galaxies. The variation of inclination angle and isophote
position angle with radius are more reliable indicators of bar
presence than the surface brightness profiles. The systematic biases
in the central $\sim 1\kpc$ of galaxies are not large. Each effect
separately gives typically a few \kms error, but the effects add up. In
some cases the error in circular velocity was a factor of two, but
typically we get about 20 percent effect.  The result is the false
inference that the density profile of the halo flattens in the central
parts.  Our observations of  real galaxies show that for a large
fraction of galaxies the velocity of gas rotation (as measured by
emission lines) is very close to the rotation of stellar component (as
measured by absorption lines). This implies that the systematic effects
discussed in this paper are also applicable both for the stars and
emission-line gas.
\end{abstract}

\section{Introduction} 

The cosmological cold dark matter (CDM) model and its present variant
 with cosmological constant ($\Lambda$CDM) have been very successful
 in accounting for the properties of galaxies and their distribution
 in space. A current problem for CDM lies in observations of dwarf
 and low surface brightness galaxies.  The rotation curves and derived
 density profiles of these galaxies appear to be in conflict with the
 CDM model \citep{M94, FP94, Burkert95, HG98}.  The problem arises
 because numerical simulations of the formation of halos in the CDM
 model reveal steeply cusped density profiles while observations favor
 shallow (core) profiles. The theory predicts that the mass density in
 the inner parts of the dark matter halos is best described as an
 $r^{-\alpha}$, $\alpha=1$ power law \citep{NFW97,Power2003,Colin2004}
 or possibly by a power-law with a slightly steeper $\alpha=1.2-1.5$
 slope \citep{Moore1999, Ghigna2000, Klypin2001, Reed03, Tasitsiomi2004}. The
 observed rotation curves of a significant fraction of dwarf galaxies,
 however, suggest that their cores do not have steep density cusps
 \citep{dBM97, BAC01, dBMBR01, MRdB01, Bolato2002, WdBW02, Simon2003}.
 Note that this is not true for some galaxies and that some
 observational results are still consistent with cosmological
 predictions \citep{vdB99, SMT2000, vdBS01}.  \citet{MRdB01} have
 analyzed H$\alpha$ observations of 50 low surface brightness galaxies
 of which 36 are of sufficient quality that rotation curves can be
 derived. The spatial resolution is improved by an order of magnitude
 over old 21~cm data.  In a companion paper \ \citet{dBMBR01}\
 conclude that at small (0.5-1~kpc) radii, the mass density
 distribution is dominated by a nearly constant density core.  The
 distribution of inner slopes is strongly peaked around $\alpha= -0.2$
 with clear contradiction with the CDM predictions.
Thus, there appears a controversy concerning the density distribution
of matter in the inner parts of dwarf galaxies. 

We propose a somewhat different approach to this problem which we believe
can help to clarify these debates.  We construct realistic models of
galaxies and ``observe'' them as real galaxies using the same
techniques or algorithms that observers apply to real data.  We study
the details of the analysis of observations and we find discrepancies
between recovered and true properties of the simulated galaxies.  We
investigate projection effects and the effects of bars and bulges on
the observed rotation curves. We show that these effects are often far
from negligible.

Some of the effects which we study, have already been discussed.
\citet{Lia1984} discussed the biases in the observed rotation curves
of high surface brightness galaxies created by strong bars.  More
recently, \citet{KW02} have investigated the effects of a weak bar on
mass determination in the central parts of spiral galaxies. They
examine rotation curves of edge-on model galaxies observed from
different viewing angles. The models had a two-dimensional solid
rotating barred potential. Motion of gas in this potential was
followed using an SPH code.  \citet{KW02} find that the mass estimated
from rotation curves is larger than the true mass by a factor of 5 for
15\% of the viewing angles. Overestimation in mass occurs more
frequently than underestimation.  \citet{KW02} focus on rotation
curves defined as the highest velocity envelope of position-velocity
diagrams.  When the rotation curve is defined as the intensity
weighted mean of position-velocity diagrams, the rotation velocity is
underestimated for most of the view angles.  \citet{KW02} conclusions
are only for edge-on galaxies; the situation with less inclined
galaxies could be more complex. Another important aspect studied by
\citet{WMN02} and \citet{WK01} is the gravity driven turbulence
induced by a weak bar in a clumpy and filamentary gaseous media. If
was found that the velocity dispersion of a cold neutral gas as well
as of an ionized component is significantly lager than naively
expected if the distribution is clumpy \citep{WMN02} or filamentary
\citep{WK01}. These results challenge the common expectation that the
gas component in a galaxy is a cold medium with low velocity
dispersion.

The effect of bulges on the observed kinematics of galaxies has its
own history \citep[e.g.,][]{Fillmore86, Kormedy89, Bertola1995,
Beltran2001,Funes2002,Pizzella2004}.  An interesting issue here is the
difference between the rotation of stars and rotation of gas in the
central 1-2~kpc region.  Because our simulations have only the stellar
component (no gas), it is important to know how different are rotation
of gas and of stars. Naively, one expects that gas as a ``colder''
component should rotate faster.  The reality is more complicated. A
detailed discussion of this issue is given in \S 2.  We also
use our own observations to measure rotation velocities in stellar and
gas components. To summarize, like other workers in this field, we find
that often in the central region the rotation of gas (as measured by
$H\alpha$ and $NII$ emission) is not very different from the rotation of
the stellar component.

Some other systematic effects were considered by \citet{Swaters2003}
who discuss effects of seeing, finite slit width, and slit
misalignment for ideal thin exponential disks. It is found that all
effects systematically underestimate the central slope. The magnitude
of the bias depends on the distance to the galaxy and on slit
misalignment.  \citet{dB03} considered the effects of an offset
between optical and kinematical centers, as well as a very simplistic
attempt to quantify the effects of non-circular motions.  Both effects
tend to lower the estimated density profile slope.
\citet{Hayashi2003} discuss effects of triaxial halos on the rotation
of gas in CDM simulations, which lead to non-circular motions.

This paper is organized as follows.  In \S 2 we present results of our
observations of rotation velocities as measured by gas and stars. In
\S 3 we discuss different observations, focusing on systematic biases,
which may affect rotation curves.  We also present results of our
observations of rotation of gas and stars. In \S 4 we present and
describe our models.  Simulations are discussed in \S 5.  We have
constructed four detailed galaxy models and run them as N-body
simulations.  In \S 6 we present our methods of analysis of the models
and in \S 7 we show our results. In \S 8 we present our conclusions.

\section{Observational measurements of rotation curves: gas vs stars} 

Most observational results which address the issue of cusps and
cores, provide rotation for gas as measured by either $H\alpha$
emission or by 21~cm line. The rotation of the stellar component is
not studied because it is difficult to measure for dwarf and LSB
galaxies. Nevertheless, there are observational results for stellar
rotation curves for a substantial number of galaxies
\citep[e.g.,][]{Fillmore86, Kormedy89, Bertola1995,
Beltran2001,Funes2002,Pizzella2004}. The observed set of galaxies cover a
wide range of morphological types (from S0's to Sc's) and
magnitudes. Most of the galaxies are normal high surface brightness
galaxies, but some are similar to dwarfs studied for the ``cusps and
cores'' problem. 
 
Let's start with what one naively should expect. Close to the center
stars are expected to have relatively large random (non-circular)
velocities. For example, if the stars are in a disk with a constant
height, the velocities perpendicular to the plane of the disk should
increase exponentially as one gets close to the center in order to
react to the exponentially increasing disk density. Other random
velocity components are coupled with the vertical velocities with the
radial rms velocity being typically twice larger than the vertical
velocity.  Non-circular stellar velocities may be even larger if there
is a bulge. Large non-circular
velocities result in significant asymmetric drift. In turn, this would
imply that rotational stellar velocity falls significantly below the
circular velocity. In the case of gas, one naively expects that
 rotation is faster because gas cools and settles in the
 plane of the disk. 

The problem with this picture, which is shared by many astronomers, is
that it is too naive. It has some justification in the outer parts of
disk galaxies.  Indeed, in our own galaxy at the solar
neighborhood molecular gas has a smaller height than 
stars. Thus, gas has smaller non-circular velocities and must rotate
faster. Young stars, which have formed relatively recently from the
gas, have smaller asymmetric drift than the older stars. This picture
of a cold thin gaseous disk inside a thicker slower rotating disk or a
bulge is naively applied to the central parts galaxies. Unfortunately,
close to the galactic center the naive picture has little
justification.  For example, the disk cannot be considered as thin
because at distances 0.3-0.5~kpc from the center the disk thickness is
comparable with the distance to the center. Gas motions also increase
substantially \citep{Beltran2001,Funes2002,Pizzella2004}

Rotation curves of ionized gas (H$\alpha$) as well as rotation and
velocity dispersion curves for the stellar component (absorption
lines) were studied by \citet{Fillmore86}. It was found that for some
galaxies the stellar component rotates {\it faster} than the ionized
gas in the central region.  For other galaxies the rotation of both
components is comparable. In particular, for the Sab galaxy NGC 4569,
it was found that at 300~pc the ionized gas rotates with a speed of
$\sim$50~km/s, while the rotation detected in the stellar component is
$\sim$100~km/s and the stellar velocity dispersion is
$\sim$150~km/s. The situation is similar inside 500~pc.  This led
\citet{Fillmore86} to suggest that gas rotation might not represent
the true circular velocity.  NGC 5879, an Sb galaxy, shows the same
rotation speed for ionized gas and for the stellar component up to the
distance of 1~kpc. However, in addition to the rotation velocity the
stellar component has a random component.  At 200~pc both stars and
ionized gas have a rotation speed of 40~km/s, but the stellar
component also shows a velocity dispersion of 70~km/s. Based on these
observations as well as on their mass models, \citet{Fillmore86}
suggest that ionized gas in these galaxies is likely to be supported
not only by rotation but also by pressure.

For NGC~4594 (M104) \citet{Kormedy89} measured the rotation of ionized
gas (HII regions) and inter-arm gas using H$\alpha$ observations. The
rotation of the stellar component was measured using absorption
lines. It was found that in the central region ionized gas rotates
with a 200~km/s, while the stellar component rotates faster: its
rotation velocity is about 300~km/s.  \citet{Kormedy89} suggest that
ionized gas in M104 has smaller rotation because it is either shed by
the bulge stars or is somehow related with the x-ray halo.  The
situation for galaxies with small bulges is not clear.  Gas in
small (few hundred parsec) bulges with an effective radius comparable
to the disk characteristic height is probably dense enough to harbor
star formation.  A coupling between young stars and ionized gas 
could produce non-circular motions in the gas
kinematics and lead to an underestimation of the central mass
content. Other possibilities are discussed and explored by
\citet{Cinzano1999}. Among these, is a
model where gas is made of clouds.  A similar situation was discussed
also by \citet{WK01} and \citet{WMN02}. The clumpy and filamentary
structure of a gaseous component considerably increases the asymmetric
drift too.  \citet{WMN02} discuss the case of NGC2915, a blue compact
dwarf galaxy, that shows a bar and a large velocity dispersion in the
H-I kinematics.

There are no strict rules as far as rotation of gas and stars in
the central parts of galaxies are concerned. In some cases the gas rotates
faster as naively expected (e.g., Sb galaxies NGC772 and NGC 7782
\citep{Pignatelli2001}). In many cases, the  rotation of gas and stars in
the central 1-2~kpc is nearly the same (e.g., Sc NGC 5530 and Sb NGC
615 \citep{Pizzella2004}). Nevertheless, there is a general trend that
late spirals with small bulges have gaseous and stellar components
rotating with almost the same velocities \citep[e.g.,][]{Cinzano1999,
Beltran2001}.

The observational data used in our paper are from a program aimed at
understanding line strengths and line strength gradients in bulges.
Long-slit spectra were obtained along the major axes of 37 spiral
galaxies using the Double Imaging Spectrograph (DIS) on the ARC 3.5m
telescope at Apache Point Observatory. The spectra cover a large
wavelength range (4000 to 7500 \AA~ or larger) at the cost of spectral
resolution (6-8 \AA~ per pixel).  The galaxies span a wide range of
inclination angles and Hubble types.  Some of the inclinations are too
low to measure rotation and the condition that H$\alpha$ is present in
emission eliminates several, mostly early-type, galaxies.  From the
sample galaxies with different morphological types and inclinations we
selected all 8 galaxies, that had inclinations in the range 45-80
degrees and had $H\alpha$ in emission at the very center and through
the most of galaxy.  We did not specifically target late type
galaxies, but, as one may expect, the condition that $H\alpha$ is in
emission at the galactic center eliminates most early types.

Spectra extracted at different positions along the slit were
cross-correlated with the spectrum at the galaxy center.  For stellar
rotation, the cross-correlation was performed within the wavelength
range 4100 to 5400 \AA~ after masking out emission features and sky
lines.  For 6 galaxies, rotation in the gas component was measured by
cross correlating within the wavelength ranges 6550 to 6600 \AA~ to
include H$\alpha$ and NII $\lambda$6583.  For the remaining two
galaxies (NGC 6368 and IC 1029) we used SII doublet (6548-6583 \AA~) because it
provided higher signal-to-noise ratio than the $H\alpha$.  A detailed
description of the observations and data analysis will be presented in
an upcoming paper (Moorthy \& Holtzman, in preparation).  Table 1
gives properties of these galaxies.

\begin{figure}[htb!]
\epsscale{0.99}
\plotone{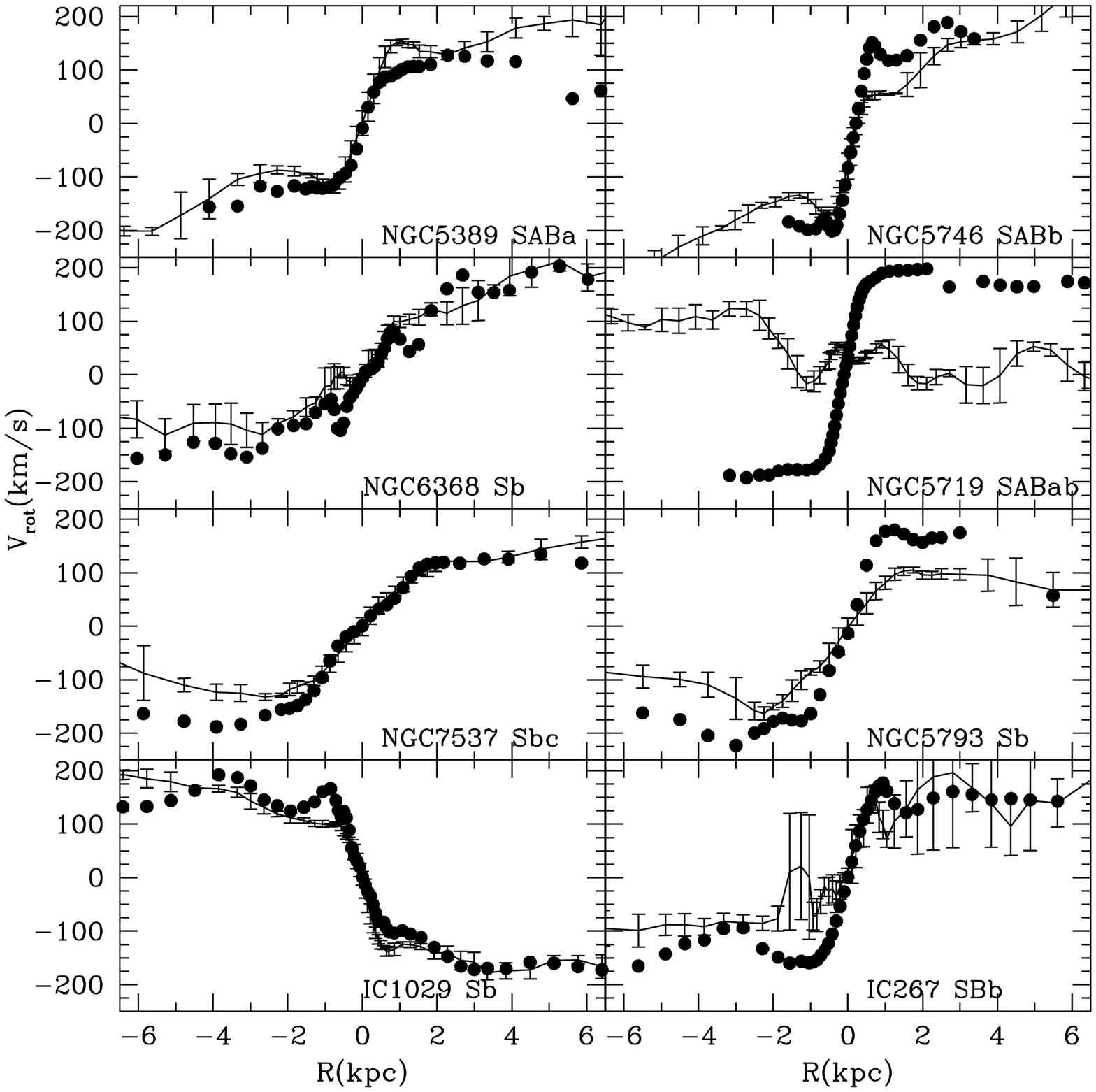}
\caption{Stellar and gas rotation velocities of observed
galaxies. Symbols show velocities measured using emission lines.  The
full curves with error bars are for absorption lines. Galaxies on the
left panels have similar rotation in gas and in stars. For galaxies on
the right gas rotates typically faster than stars. Of those three
galaxies are barred.}
\label{fig:rotgalax}  
\end{figure}

\begin{figure}[htb!] 
\epsscale{0.99}
\plotone{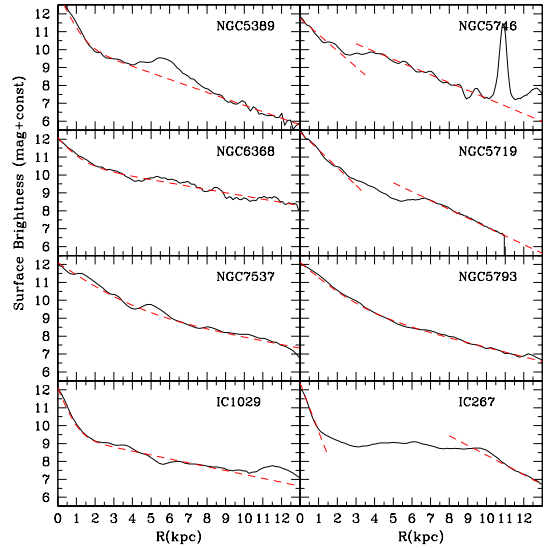}
\caption{Surface brightness profiles. Dashed curves show disk/bulge
double exponential approximations. In the case of barred galaxies the
exponentials do not join though the outer disk are still clearly
exponential. The spike at 10~kpc in profile of NGC5746 is due to a
star. The ring of NGC5389 is seen at 6kpc as a hump overlaid on exponential disk.}
\label{fig:sbgalax}  
\end{figure}

Rotation curves of the galaxies are shown in 
Figure~\ref{fig:rotgalax}.  Surface brightness profiles along the
spectrograph slit are shown in Figure~\ref{fig:sbgalax}.  Both the
velocities and the surface brightnesses are measured along the major
axis of the galaxies.  For barred galaxies they are long the bars. A
large fraction (but not all) galaxies in our sample show stellar
rotation close to that of the gas. All three barred galaxy in our
sample, have complicated velocity patterns. In case of NGC5719 there
is little stellar rotation along the bar. Yet, the bar in the galaxy
is quite extreme: more than 10~kpc. This galaxy  also presents a 
peculiar morphology with a single strong dust lane oriented in different 
direction with respect to the galaxy major axis.  
A large bulge dominates the central light distribution. 
Based on spectroscopic observations in visible, near and far infrared light  
as well as in radio wavelengths \citet{HHSH} and \citet{HHRH} conclude that NGC5719 is  
a composite of an AGN and a starburst galaxy.     
This evidence combined with the large size of the bar and the existence of a companion  
NGC 5713, suggests that  NGC5719 is suffering a strong interaction. 
\citet{Hunter02} recently presented the case    
of NGC4449. This galaxy is also strongly barred and shows no rotation in  
the stellar component. \citet{Hunter02} propose a model where the orientation  
of a gaseous younger disk and the stellar component are different as a result   
of an interaction.  The model introduced by \citet{Hunter02} supports our 
interpretation of NGC5719, however a detailed modeling of this peculiar
galaxy is beyond the scope of this paper. The other barred galaxies in our 
sample show large and strong bars too.

We clearly find a tendency that stars in galaxies with smaller bulges 
rotate as fast as the gas. This is compatible with previous results 
\citep[e.g.,][]{Cinzano1999,Beltran2001}.  There are important 
conclusions, which we get form these results.  (1) Because the stars and the 
gas move in the same gravitational potential, observed $H\alpha$ 
rotation curves must experience the same systematic biases as the 
stellar rotation curves.  (2) We can use stellar-dynamic simulations to 
study those effects. (3) The environment of a galaxy should be taken in account
in the interpretation of the kinematics, specifically for the case of 
strongly interacting galaxies.

\section{Observational measurements of rotation curves: systematic effects}

For most galaxies corrections for finite disk thickness seems to be
quite small. Indeed, at few kiloparsec distance from the galactic
center corrections for disk height of few hundred parsec are not
significant.  As far as we know, no correction for finite disk
thickness was ever made.  Is this justified for central regions
(0.5-1\kpc) of dwarf galaxies? It is far from obvious. Observations of
edge-on galaxies indicate that as the function of decreasing circular
velocity the disk height decreases slower than the exponential disk
scale length \citep{Kregel2002, Dalcanton2003}. In other words, dwarf
galaxies are relatively thicker than giants. Using I-band photometry
for 34 edge-on spiral galaxies \citet{Kregel2002} give estimates of
the exponential heights.  For 6 galaxies with maximum circular
velocity below $100~\kms$ their results give the average height of
440~pc.  One of the thinnest galaxies is UGC 7321 \citep{Matthews2000}
with an exponential height of 150~pc, which is about what we use for
our model IV ``Disk''.  Interestingly enough \citet{MatthewsWood2003}
find that the neutral hydrogen has about the same height as the
stellar component. If the galaxies, which are used to study rotation
curves and the dark matter density have the same heights as those
observed edge-on, it is clear that in the central sub-kiloparsec
region, the disks cannot be treated as thin at all.

\begin{figure}[htb!]
\epsscale{0.8}
\plotone{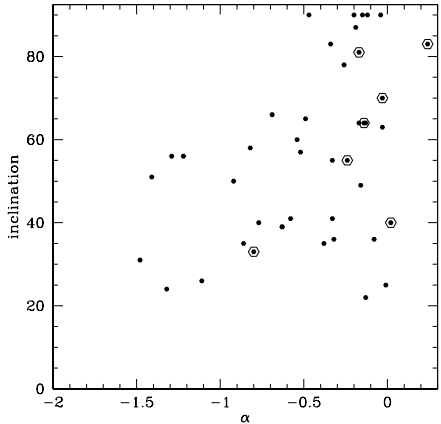}
\caption{Inclination angle versus inner power law slope $\alpha$ for
the galaxies in the \citet{dBMBR01} sample.  The circled points
correspond to galaxies that have been identified by \citet{MRdB01} as
being barred galaxies. There should be no correlation of slopes with
the inclination. Lack of slopes at left top corner clearly indicates
significant biases in the sample.}
\label{fig:deblok}  
\end{figure}

The effect of finite disk thickness increases with increasing galaxy
inclination.  As part of our study we have examined the data of
\citet{MRdB01} and \citet{dBMBR01} to see if we can find evidence of
observational effects involving inclination and galaxy morphology.
\citet{MRdB01} list the inclinations for a sample of 50 galaxies which
they observed in H$\alpha$.  \citet{dBMBR01} list values for the inner
power law slope ($\alpha$) of $\rho(r)$ for 48 galaxies. There is a
partial overlap with the sample of \citet{dBMR01}, other galaxies
being culled from data published by \citet{dBB02}, \citet{SMT2000} and
\citet{V97}. We have combined these data to produce a list of 43
galaxies. These data are shown in Figure~\ref{fig:deblok}. If the
estimates of the inner density slope $\alpha$ had no biases, there
would be no dependence of $\alpha$ with the inclination. This clearly
is not so: the top left corner of the plot does not have points.

Also note that galaxies identified as barred
galaxies tend to have larger values of $\alpha$ than the average. All
but one have $\alpha > -0.25$.  We do not have morphological
information for all galaxies in the sample. The data in
Figure~\ref{fig:deblok} suggest that galaxies with high inclinations
have large $\alpha$ values. Galaxies with inclinations $\gtrsim
60$\arcdeg\ have $\alpha$ values greater than -1.  The evidence
suggests that projection effects and the presence of bars influences
must be taken into account when determining density distributions from
rotation curves.

Recent observations of galaxies suggest that  
galactic bars occur more frequently than had been  
previously suspected. \citet{Esk02} have presented BVRJHK imaging of a
sample of 205 bright nearby spiral galaxies. An examination of Table I
in their paper reveals that in the B-band 103 of the galaxies are
classified as barred spirals whereas in the H band 140 or $\sim$70\%
of the sample are classified as barred spirals.  The conclusions we
draw concerning the effects of bars on the interpretation of spiral
galaxy rotation curves may well be relevant to the majority of spiral
galaxies.

Dwarf and LSB galaxies used for analysis of rotation curves and the
dark matter content typically are selected to be ``bulge-less'': types
Sc-Sd. Yet, it appears that many of them still have small bulges.
\citet{BSM03} have obtained HST WFPC2 images of a sample of 19 nearby
spiral galaxies of late Hubble type. For these galaxies bulges are not
evident in the ground based optical data for these galaxies.  The HST
images were combined with the existing ground based optical images to
obtain surface brightness profiles.  The conventional wisdom has it
that the sequence from early to late-type galaxies ranges from bulge
dominated to disk dominated galaxies. The reality seems to be that
only $\sim$30\% of the galaxies in the sample are pure disk galaxies
\citep{BSM03}.  The remainder are well fitted by adding a bulge
component.  Some of these bulges are small, extending out to $\sim$
300~pc.  Larger bulges extend out to 2~kpc.  As in the case of bars, the
evidence suggests that the presence of bulges in spiral galaxies must
be taken into account in the investigation of rotation curves of
spiral galaxies.  \citet{GDIT02} confirm these observations in a near
infrared study of 88 low surface brightness galaxies and normal
spirals. They note that many of the low surface brightness galaxies
classified as late types based on optical B-band images show a
dramatic bulge in the near-IR. The data support the conclusion of
\citet{BBH99} that many low surface brightness galaxies could be
regarded as high surface brightness bulges embedded in low surface
brightness disks. Since the surface brightness in the near-IR closely
reflects the stellar surface density it is important to take bulges
into account when modeling the rotation curves of low surface
brightness galaxies. 

Note that the bulges are really small. A 300~pc
bulge is hardly thicker than the disk. Physical conditions in such
small bulges are not much different from those in the disk.
Stars can form in those bulges.

The next important point to keep in mind is that of accuracy.
How accurately can one determine the inclination of a galaxy?
We refer to NCG 3109, a Magellanic type spiral that we have attempted
to model in some detail \citep{VRK03}. \citet{JC90} quote several values
of inclination for this galaxy of $80\arcdeg\pm 2\arcdeg$ based on B-band 
photometry.  The same authors find a value of  $75\arcdeg \pm 2\arcdeg$ based
on photometry of an I-band plate taken using the U.K. Schmidt 
telescope. In a more recent study, \citet{BAC01} present CFHT Fabry-Perot
H-$\alpha$ observations of NCG 3109. These data show an inclination
between $85\arcdeg$ and $90\arcdeg$ with one inner point at $74\arcdeg$.
Although the formal uncertainty in inclination
in a given waveband may be as small as 2$\arcdeg$, a more realistic value 
is at least $5\arcdeg$ for the uncertainty in a galaxy's inclination.
\citet{BS02}) support this conclusion. Based on 
a study of 74 spiral galaxies they find that systematic uncertainties 
in position angle and inclination due to 
the presence of nonaxisymmetric structure are of order $\sim 5\arcdeg$.
These estimates are of course lower limits on the real errors.
\citet{WdBW02} publish inclinations and position angles of the local group 
dwarf NGC 6822 as a function of radius 
based on radio data obtained with the Australia Telescope Compact
Array. The values for the inclination angle of this galaxy as 
a function of radius fluctuate between $\sim 80\arcdeg$  and $20\arcdeg$!
In their analysis, a constant value $60\arcdeg$ is adopted without 
further explanation. The position angle for this galaxy 
varies between $100 \arcdeg$ and $140 \arcdeg$ according to these 
data.

To summarize, there are three key points that have guided our
analysis. Firstly, the central density profile obtained for a given
galaxy appears to be influenced by the inclination of the galaxy,
for inclinations greater than $60\arcdeg$ at least (it is 
difficult to determine the inclination of these dwarf galaxies 
to better than several degrees; $\sim 5\arcdeg$).
Secondly, the 
data seem to indicate that galaxies with bars have densities 
that flatten in their centers when the standard analysis is performed.
Errors of several degrees ($\sim 5\arcdeg$) can occur when 
determining inclinations.
Thirdly, given increasing evidence for bulges in spiral galaxies of all
morphological types, it is important to examine the biases that 
bulges (as well as bars) introduce into the determination of 
central density profiles.

\section{Models}

We use $N$-body simulations to produce models of galaxies. The models
include realistic dark matter and stellar component, but do not have
gas.  Gas would significantly complicate numerical models making
predictions less certain.  In dwarf galaxies gas is very patchy and
often has significant non-circular motions
\citep[e.g.,][]{Bolato2002,Simon2003}.  The stellar component has its own
complications as we find below, but those are relatively simple and
easy to interpret as compared with the gas motions. In dwarf
galaxies, the gas has all the complications of stars and in addition it brings
its own problems. Studying the stellar component is definitely a good
starting point.

We study the kinematics and spatial distribution of stellar component in
four models.  Parameters of the models are given in Table 2. All the
models are scaled to have a maximum circular velocity of approximately
60~\kms, which is the value for dwarf galaxies such as NGC3109
\citep{JC90} and NGC6822 \citep{WdBW02}. The virial masses for these
models are in the range $(4-7)\times 10^{10}\Msun$. In all cases the
baryonic component represents a very small fraction (1\%-2\%) of the
virial mass, which is a factor of ten smaller than the cosmological
ratio of the baryons to the dark matter.

All models have extended ($\sim 100~\kpc$) halos, which have high
concentration NFW density profiles. We did not try to make more
realistic fits for the dark matter halos. In reality the halos should
be slightly perturbed in the central part because of the presence of
baryons.  We decided to neglect this effect in favor of simplicity of
the models. Most of our models have very thin disks. Real disks are
typically thicker than what we assume for our models. The only model,
which has a realistic disk is the ``Disk'' model V.

The first two models (dubbed ``Bar'' and ``Dwarf'') start with a pure
stellar disk embedded in an extended dark matter halo. Initially the
disks in the models are submaximal: about 50\% of mass in central
1~kpc is in the disk the rest is in the dark matter. The disks are
relatively cold (stability parameter Q is 1.2). As the result of this
combination of parameters, the disks are unstable and produce
bars. The main difference between the models is that the disk in the
model ``Bar'' is significantly more massive than in the model
``Dwarf''.

Models ``Bulge'', ``Thin Disk'', and ``Disk'' do not have bars. The models are stable
- they do not evolve much.  Nevertheless, we run the models for about
1.5-2~Gyrs to ensure detailed equilibrium.  Indeed, small variations
are found in the models. Initially the rotational velocity does not depend
on the distance $z$ to the disk plane. After evolution we find that
the rotational decreases with $z$. This tendency increases effects of galaxy inclination.
Models ``Thin/Thick Disk'' have relatively hot
disks ($Q=3.0$).  The disks are also more extended, which make them
sub-dominant even in the center. The disks did not develop a bar.

\begin{figure}[htb!]
\epsscale{1.00}
\plotone{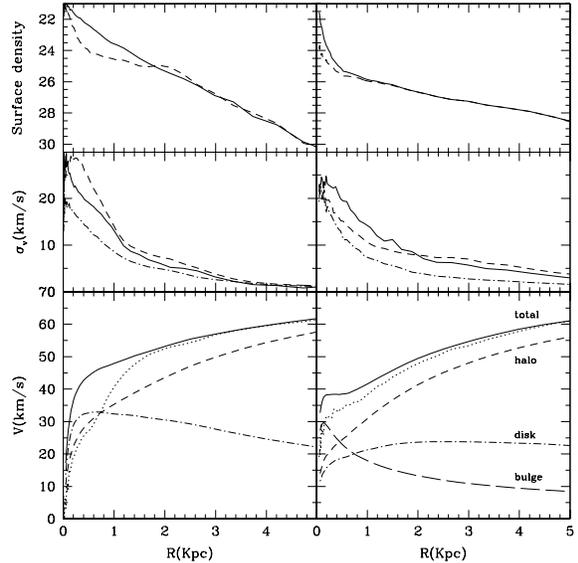}
\caption{Profiles for the final moments of the model II (``Dwarf'',
left panels) and the model III (``Bulge'', right panels).  Lower
panels show circular velocities of different components in the
models. The dotted curves show the rotation velocity of the disk
components. For both models the baryonic component (disk+bulge)
dominates inside the central 700~pc radius.  Middle panels show
velocity dispersions for the baryonic components. The dot-dashed
curves are for the vertical (z) velocity. The full and dashed curves
are for tangential (in the plain of the disk) and radial velocities
correspondingly. Note that velocities are very small. Only inside
central 500~pc do they go above $20~\kms$. The top panels present the
surface density profiles of the baryonic components scaled to
magnitude units with arbitrary zero points. For the barred ``Dwarf''
the full curve is for the surface brightness along the bar and the
dashed curve is perpendicular to the bar.  The length of the bar is
$1.8~\kpc$. The full curve for the ``Bulge'' model shows the total
disk+bulge surface brightness. Disk component is shown by the dashed
curve.}
\label{fig:models}
\end{figure}

Figure~\ref{fig:models} presents details of the distribution of different
quantities for models II and III.  The models to some degree are
similar and are representative for the other two models (I and IV).  In
 models II and III  the dark matter is dominant in the outer regions, but is
sub-dominant in the central 500~pc.  Velocity dispersions of baryonic
components are very small. For example, for models II and III the rms
velocities are smaller than $15~\kms$ even at $1\kpc$ radius. 

\section{The Numerical Simulations}

The procedure of generating of initial conditions is described in detail in
\citet{ValenzuelaKlypin2003}.  We use the following approximation for
the density of the stellar disk in cylindrical coordinates:
\begin{equation}
 \rho_d(R,z) =\frac{M_d}{4\pi z_0R_d^2}e^{-\frac{R}{R_d}}
                       sech^2(z/z_0),
\label{eq:expdisk}
\end{equation}
where $M_d$ is the mass of the disk, $R_d$ is the exponential length,
and $z_0$ is the scale height. The later was assumed to be constant
through the disk. Effective exponential scale height (height at which
the density decreases by $e-$times) is equal to $z_e =1.085 z_0$.
The disk is truncated at five exponential lengths
and at three scale heights: $R<5R_d$, $z<3z_0$. Disk mass was slightly
corrected to include the effects of truncation. The vertical velocity
dispersion $\sigma_z$ is related to the surface stellar density
$\Sigma$ and the disk height $z_0$:
\begin{equation}
 \sigma^2_z(R) =\pi G z_0 \Sigma(R),
\label{eq:vertical}
\end{equation}
where $G$ is the gravitational constant. The radial velocity
dispersion $\sigma_R$ is also assumed to be directly related to the
surface density:
\begin{equation}
 \sigma^2_R(R) =A e^{-\frac{R}{R_d}}.
\label{eq:radial}
\end{equation}
The normalization constant $A$ in eq.~(\ref{eq:radial}) is fixed in
such a way that at any radius $R$ the rms radial
random velocity is $Q$ times the critical value needed to stabilize
a differentially rotating disk against local perturbations.
The tangential ($\phi$) component of the rotational velocity and its
dispersion are found using the asymmetric drift and the epicyclic
approximations. For the disk component we use the thin-disk
approximation.

For model ``Bulge'' we also add a spherical non-rotating component
with the Hernquist density profile:
\begin{equation}
 \rho_{\rm dm}(r) = \frac{\rho_s}{x(1+x)^3}, \ x=r/r_H, 
\label{eq:Hernq}
\end{equation} 
where $r_H$ is the 1/4-mass radius. For the model $r_H=125~\pc$, which gives half-mass radius 250~pc.

We assume that the dark matter density profile is described by the NFW
profile \citep{NFW97}:
\begin{eqnarray}
 \rho_{\rm dm}(r) &=& \frac{\rho_s}{x(1+x)^2}, \ x=r/r_s, \\
 M_{\rm vir} &=& 4\pi\rho_s r^3\left[ \ln(1+C) -{C\over 1+C}\right], \\
   C&=&{r_{\rm vir}\over r_s},
\label{eq:NFW}
\end{eqnarray} 
where $M_{\rm vir}$ and $C$ are the virial mass and the concentration
of the halo.  For given virial mass the virial radius of the halo is
found assuming a flat cosmological model with matter density parameter
$\Omega_0=0.3$ and the Hubble constant $H_0=70\kms {\rm Mpc}^{-1}$.

Knowing the mass distribution of the system $M(r)$, we find the radial
velocity dispersion of the dark matter:
\begin{equation}
 \sigma_{r, {\rm dm}}^2 ={1\over \rho_{\rm dm}}\int_r^{\infty}\rho_{\rm dm}\frac{GM(r)}{r^2}dr.
\label{eq:radialDM}.
\end{equation}
The other two components of the velocity dispersion are equal to
$\sigma_{r, {\rm dm}}^2$. In other words, the velocity distribution is
isotropic, which is a good approximation for the central parts of the
dark matter halos.  At each radius we find the escape velocity and
remove particles moving faster than the escape velocity.
The dark matter halo is truncated at the virial radius,

Velocities are picked from an appropriate Gaussian distribution
truncated at the escape velocity.  The disks are realized with
particles of an equal mass. The dark matter halo is composed of
particles of variable mass with small particles placed in the central
region and larger particles at larger distances.  We double the
particle mass as we move further and further from the center.  The
region covered by small mass particles is $\sim 20$~kpc -- much larger
than the size of the disk. In order to reduce the two-body scattering
each dark matter particle in the central region has the same mass as a
disk particle.  This procedure is designed to reduce the number of
particles and, at the same time, to allow us to cover a very large
volume.

We use the Adaptive-Refinement-Tree (ART) $N$-body code
\citep{kkk:97,kravtsov:99} to run the numerical simulations analyzed in 
this paper.  The code starts with a uniform grid, which covers the 
whole computational box. This grid defines the lowest (zeroth) level 
of resolution of the simulation.  The standard Particles-Mesh 
algorithm is used to compute the density and gravitational potential 
on the zeroth-level mesh with periodical boundary conditions.  The 
code then reaches high force resolution by refining all high density 
regions using an automated refinement algorithm.  The refinements are  
recursive. A refined region can also be refined. Each subsequent 
refinement level has half of the previous level's cell size. This creates 
a hierarchy of refinement meshes of different resolution, size, and 
geometry covering regions of interest. Because each individual cubic 
cell can be refined, the shape of the refinement mesh can be arbitrary 
and effectively match the geometry of the region of interest. This 
algorithm is well suited for simulations of a selected region within a 
large computational box, as in the simulations presented below. 

The criterion for refinement is the local density of particles. If the 
number of particles in a mesh cell (as estimated by the Cloud-In-Cell 
method) exceeds the level $n_{\rm thresh}$, the cell is split 
(``refined'') into 8 cells of the next refinement level.  The 
refinement threshold depends on the refinement level. The threshold    
for cell refinement was low on the zeroth level: $n_{\rm 
thresh}(0)=2$.  Thus, every zeroth-level cell containing two or more 
particles was refined.  The threshold was higher on deeper levels of 
refinement: $n_{\rm thresh}=3$ and $n_{\rm thresh}=4$ for the first 
level and higher levels, respectively.  

During the integration, spatial refinement is accompanied by temporal 
refinement.  Namely, each level of refinement is integrated with its 
own time step, which decreases by factor two with each refinement.  
This variable time stepping is very important for accuracy of the   
results.  As the force resolution increases, more steps are needed to    
integrate the trajectories accurately. 

\section{Analysis}
We have analyzed the N-body simulations using several methods
or algorithms that mimic those used by the observers. Our data 
consist of velocities, positions and masses of the particles that 
constitute the disk of the model galaxy. The simulations are 
set up so that the plane of the disk is the $x-y$ plane. During
our 'observations' we 'look' down the $z-$axis of the simulation.
To observe our galaxy at various inclinations we simply rotate 
$y-z$ plane about the $x-$axis by varying angles.

The object of the exercise is to calculate rotation curves for 
our model galaxies as well as reconstruct the density profile. 
We have also constructed velocity field diagrams in the form of  
velocity contour plotted on the sky. 
One method frequently used by optical observers is long 
slit spectroscopy. In our case, we place a rectangular 
slit running parallel to and centered on the x-axis. We then 
select all particles whose $x-y$ coordinates place them 
within the slit boundaries. We define pixels along the slit
and we assign particles falling in the slit to these pixels.
We compute the line-of-sight velocity of each particle. We then compute
the  average velocity of the particles in each pixel.
This average velocity, determined as a function of distance
along the slit, is our ``observed'' rotation curve. By comparing 
this average rotation velocity with the true average velocity
as well as the circular velocity, we can estimate the accuracy 
of the method.

In the case of HI observations in the radio and Fabry-Perot
measurements in the optical, one has velocity information
over the whole galaxy image not just along a narrow slit.
In this case, observers determine the rotation curve using 
a tilted ring model of the velocity field. The method has been 
described by \citet{B89}. We reproduce the salient points below

\subsection{The Tilted Ring Model}

The model is based on the assumption that a rotating disk galaxy can be 
described by a set of concentric rings. Each ring has a 
constant circular velocity V$_c$ and two orientation 
angles (the inclination angle, $i$, and the position 
angle of the major axis, $\phi$). The values of the ring 
parameters are determined from the observed 
radial velocities in a set of concentric elliptic annuli in the 
plane of the sky. The ellipticity of the annuli, determined from 
isophotal analysis, fixes the inclination of each ring. 
$V_0$ is the systemic velocity of the galaxy
(in our case 0 \kms). For a given 
ring, the observed radial velocities recorded on a set of sky
coordinates $(x,y)$ are related to the basic parameters by:
\begin{equation}
  V(x,y)=V_0+V_c(R)\sin(i)\cos(\theta).
\end{equation}
In the expression above, $\theta$ is the azimuthal angle in
 the plane of the galaxy, related to the parameters
$i, \phi, x_0$ and $y_0$ through:

\begin{eqnarray}
\cos(\theta) = { -(x-x_0)\sin(\phi)+ (y-y_0)\cos(\phi) \over R} \\
\sin(\theta) = { -(x-x_0)\cos(\phi)+ (y-y_0)\sin(\phi) \over R\cos(i)},
\end{eqnarray}
where $R$ is the mean radius of the ring in the plane of the galaxy.

The first step in such a procedure is to determine the the position
angle and ellipticity of the `isophotes' of the model galaxy image.
To do this we project all the particle positions on the sky plane.
We then form an image in FITS format
that we process using IRAF \footnote{IRAF is written and supported by the
IRAF programming group at the National Optical Astronomy Observatories
(NOAO).}. We use a package within IRAF called STSDAS \footnote{This is the software
for reducing and analyzing data from the Hubble Space Telescope. It is
layered on top of IRAF and provides general purpose tools for astronomical data
analysis as well as routines specifically designed for HST analysis.}. 
Within STSDAS we use the ELLIPSE task which to fit a set of 
elliptical isophotes over the image. The method has been 
described by \citet{J87}.

Line-of-sight velocities measured at positions close to the minor axis carry less
information about the underlying circular velocity than those measured at positions
close to the major axis. We have excluded all values of the angle 
$\theta > 45\arcdeg$ where $\theta$ is the angle with respect to the 
major axis in the plane of the galaxy. \citet{WdBW02} experimented with
different weighting schemes, such as 
down weighting the minor axis by $|\cos \theta|$ and $\cos^2 \theta$.
They also varied the exclusion by angle between 45\arcdeg and
15\arcdeg and found no difference in the curves produced for their analysis 
of the galaxy NGC 6822.

Once we know the values of $\phi$, $i$ and $R$ we can compute
$V_c$ for the each elliptical annulus and thus obtain $V_c(R)$, the 
rotation curve. As with the long slit method, we can compare $V_c(R)$
with the known rotational velocities and true circular velocities
for the model galaxy.

\subsection{Density profiles}

One can invert the observed rotation curve to determine  
the parent mass distribution \citep{dBMBR01} using the  
equation  
\begin{equation} 
4\pi G\rho(r) = {2v \over r}{\partial v \over \partial r}+ \biggl ( {v \over 
r } \biggr)^2. 
\end{equation} 
This procedure assumes that the galaxy is spherically symmetric,  
completely dark matter dominated  and that the gas moves in circular  
orbits in a planar disk.  
It is worth  checking  the reliability of the method using a realistic model  
of a galaxy.    

\section{Results}
\subsection{Long Slit Observations}
\subsubsection{Models IV and V (no bar, no bulge)}

We observe the models IV and V in a manner designed  
to mimic long slit observations. In the following discussion we take   
circular velocity to mean the velocity at which a particle would  
move if it was moving in uniform circular motion around the center of  
the galaxy. The term rotational velocity refers to the speed at which a
particle actually is rotating about the center of the galaxy; the total   
velocity of the particle can of course be larger than this. The term  
''observed velocity'' refers to the mean observed velocity at a 
given radius from the center. As we shall show, this latter number  
can have a value that is considerably smaller than the  
rotational velocity which in turn can be considerably smaller than the   
circular velocity.   

\begin{figure}[htb!]
\epsscale{1.00}
\plotone{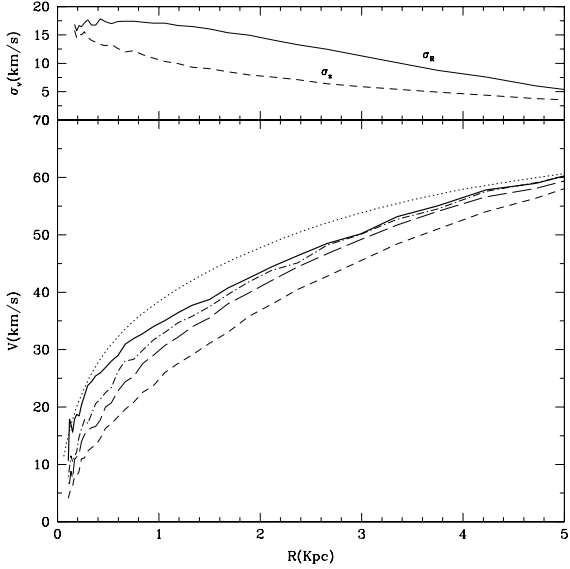}
\caption{Velocities in Model V. The bottom panel shows the true
rotation velocity (full curve) and the circular velocity (dotted
curve) for this model. Long slit observations underestimate the
rotational and circular velocities. The effect increases with
inclination angle. The dot-dashed (long and short dashed) curves are
for inclination $70\arcdeg$ ($80 \arcdeg$ and $90 \arcdeg$).  The top
panel presents rms velocities of the stellar component. The random
velocities are small (10 --15\kms), but they are responsible for the
difference between the (true) rotation and the circular velocities.}
\label{fig:thickdisk}
\end{figure}

Figure~\ref{fig:thickdisk} compares the rotation curve obtained in
this manner with true rotation velocities in the disk which we can
measure directly. We also determine the total circular velocity that
would be measured if there were no random velocities in the disk
(i.e. perfect circular motion in the galaxy potential). In order to
mimic slit observations, we incline the model by some angle and find
line-of-sight velocities for ``stars'' inside a long slit of 100~pc width.

A clear trend is that the discrepancy between the ``observed''
and true rotation curves increases with inclination of 
the galaxy. By the time one reaches an inclination 
angle of $80 \arcdeg$ the discrepancy is $\sim 8$ \kms\ at 
a radius of 0.5~kpc. This is a 30\% 
in the estimation of velocity in the central part of the galaxy.

Why does the long slit mode of observation lead to underestimates of
the rotational velocities in the disk?  Let us consider a simple
example. For model IV we take a line of sight through the plane of the
disk whose closest approach to the center of the galaxy is 0.7~kpc. We
choose this number since it is within the central kiloparsec of dwarf
galaxies that mismatch between theory and observation is
greatest. Averaging the velocities of N-body particles along this line
of sight is equivalent to measuring the rotation velocity at one pixel
location of a long slit observation.  The velocities are weighted by 
the density of material along the line of sight. Ironically, the 
smallest rotational velocity occurs for the points that are located 
closest to the center of the galaxy. Since we are averaging this 
velocity with velocities of points further from the center that thus  
have larger circular velocities, how can the average be lower? The 
point of course is that we measure radial velocities along the line of 
sight. For example, a particle located 5~kpc away from the center of 
our model rotates at 60~\kms. However, the projected velocity of this 
particle on a line of sight 0.7~kpc from the center of the model is 
only 9~\kms. The rotational velocity of the disk 0.7~kpc from the 
center is 30~\kms. The average velocity that we measure along this
line of sight is 20.4~\kms\ or 70\% of the true velocity.  One might
well expect this difference to be even larger given that we are
including radial velocities as low as 9~\kms. The answer is that the
integral is intensity weighted so that points near the center carry
considerably more weight in the mean determination than those near the
edge of the galaxy. For example, for the whole line of sight we get a
mean velocity of 20.4~\kms while we get 26~\kms if we include only the
central kpc.

\begin{figure}[htb! ]
\epsscale{1.1}
\plottwo{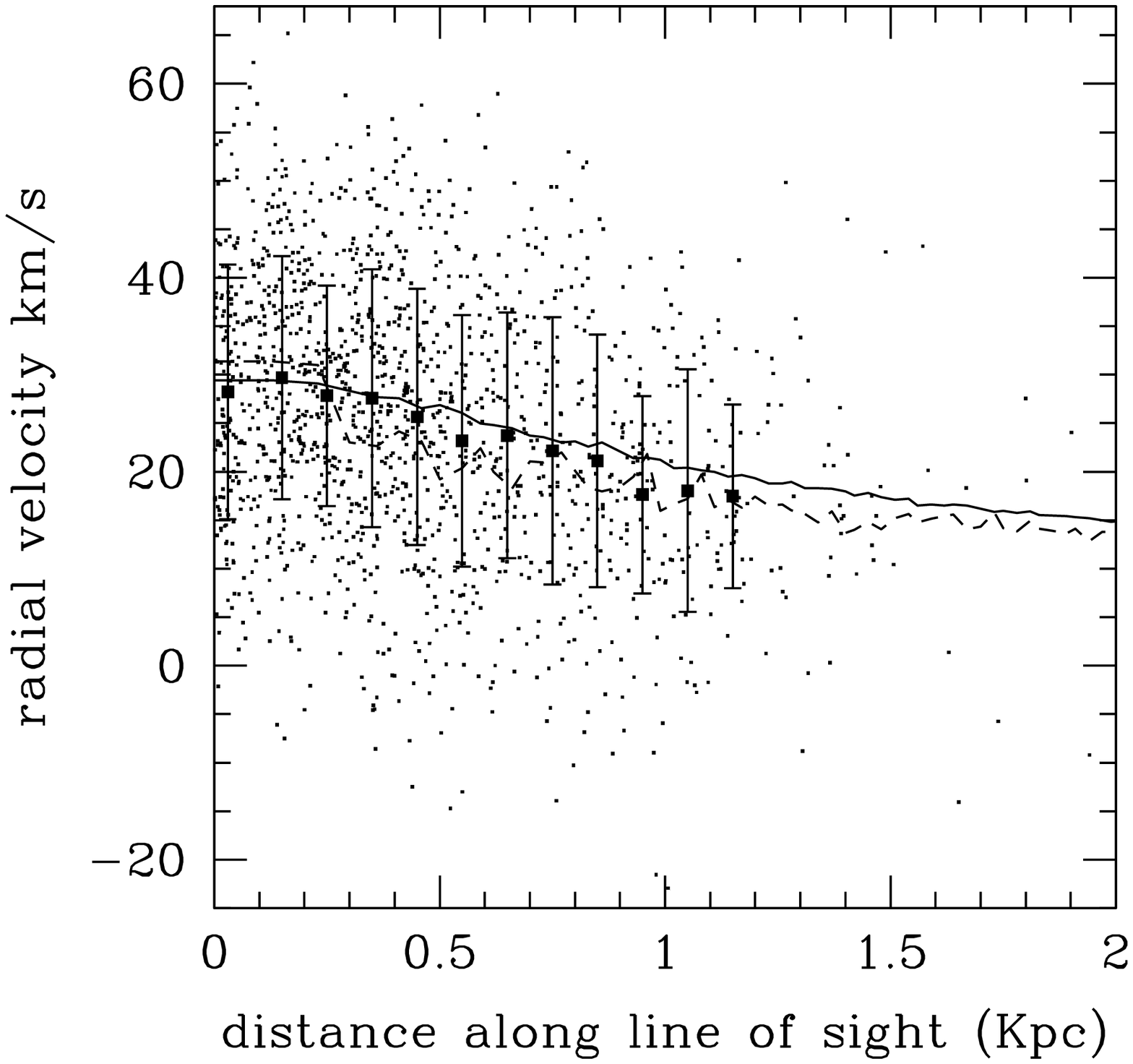}{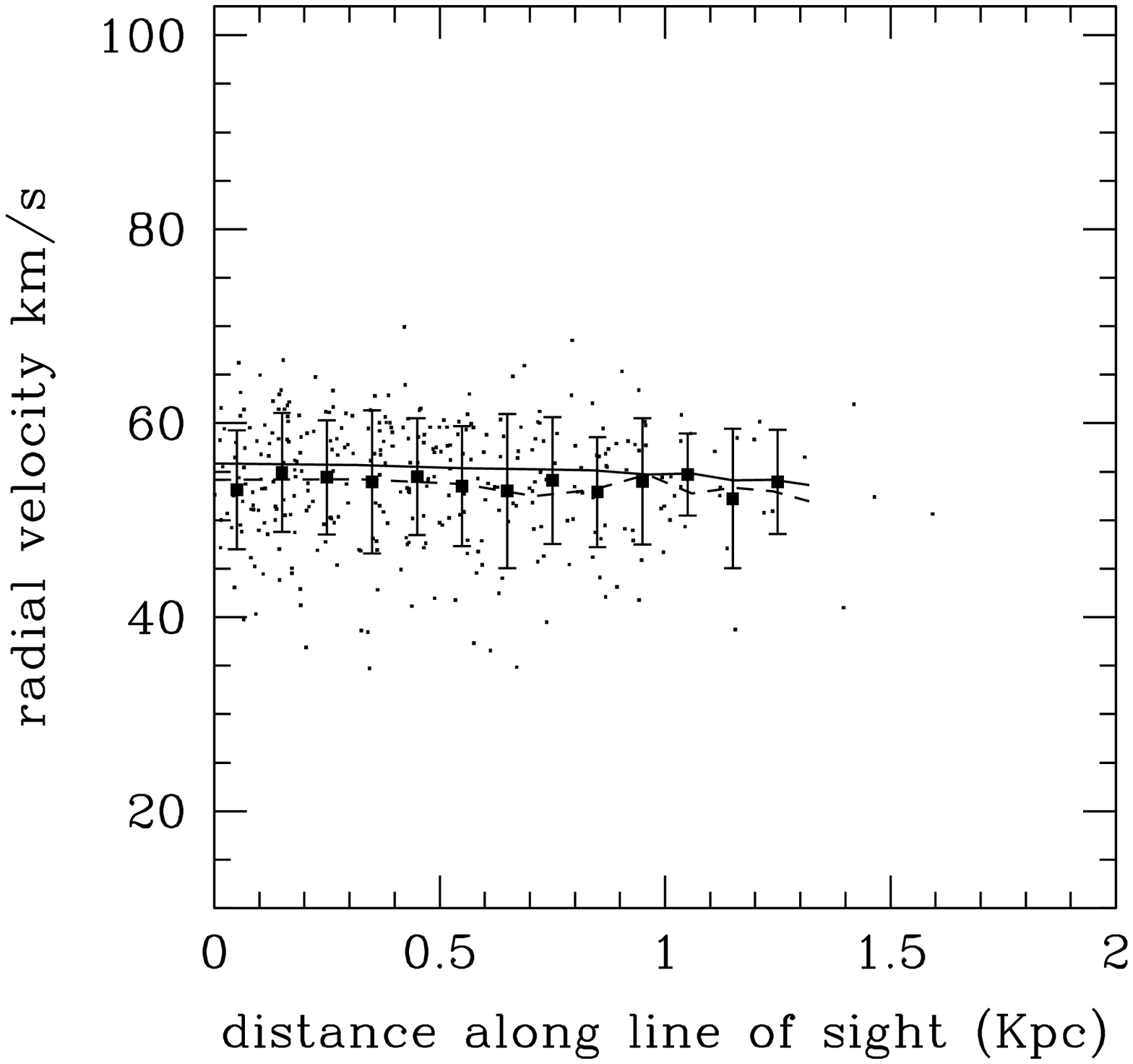}
\caption{Distribution of line-of-sight velocities versus distance  along 
the line of sight for 
particles selected at 0.6 to 0.8~kpc (left panel)  and at 4.0 to 4.2~kpc
(right panel) along a `` slit'', centered
on the major axis for Model IV ``Thin Disk''  viewed at an 
inclination of 75$\arcdeg$. 
The solid line shows the average velocity of particles within 0.25~kpc
of the galactic plane, projected along our chosen line of sight.
The dashed line shows the projected rotational velocities of particles
further than 0.25~kpc from the galactic plane. The points with error
bars show the mean velocity in bins 0.1~kpc wide.
}
\label{fig:zvz}
\end{figure}

The discussion above refers to the model viewed edge-on. The  
effect of viewing the galaxy at a lower inclination is to include 
particles that are above and below the galactic plane in the integral. 
Figure~\ref{fig:zvz} shows the radial velocity versus line of  
sight distance for particles viewed with an inclination  
angle of 75$\arcdeg$. The points in the plot on the left are selected at location 
along the slit of 0.6 to 0.8~kpc, those in the right hand plot are 
selected from 4. to 4.2~kpc. 
It is clear (as expected) that particles located  
above the plane of the galaxy have slower rotational velocities. The 
mean velocity measured as a function of distance along the line  
of sight matches the rotation curve computed for particles $0.25~$kpc above  
the disk at about $1~$kpc, which is where one would expect it  
to for an inclination angle of 75\arcdeg. 
Note that particles further than 0.25~kpc from the plane 
of the galaxy are not substantially affecting the velocity 
measurement because the density out there is too low.

\begin{figure}[htb!]
\epsscale{0.99}
\plotone{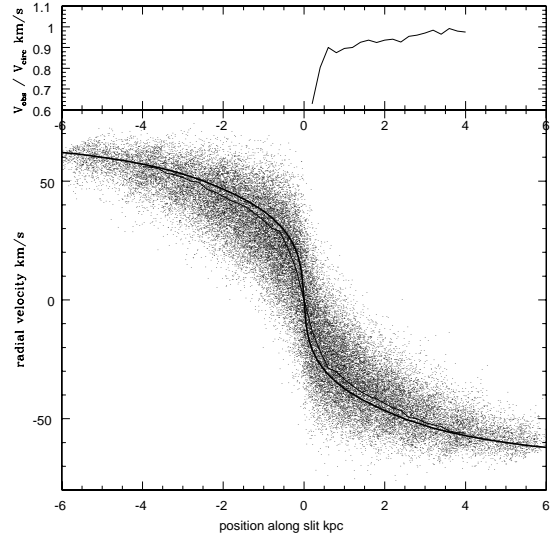}
\caption{Radial velocities along a ``slit'' for model IV inclined by
75$\arcdeg$. All velocities are corrected for the inclination. Bottom
panel: distribution of velocities of individual particles. The 
curve shows the true rotation velocity.  The top panel shows the
ratio of observed to circular velocities. The velocity is underestimated
by 20\% at 0.5~kpc, while at distances larger than 2~kpc the errors
are very small.}
\label{fig:zvz2}
\end{figure}

One might ask why, given these projection effects, the rotational 
velocity is only underestimated for lines of sight passing close to
the center of the galaxy. Lines of sight 
that pass more than $\sim$2~kpc from the  center the measured velocities
are in good agreement with the true rotational velocities for our 
models. The right hand plot of Figure~\ref{fig:zvz} illustrates 
this; we do a good job recovering the rotational velocity of the 
particles at large radii. When the distance of the line of sight from the center of 
the galaxy is comparable to the size of the disk, the projection 
effects are quite different to those discussed in the previous case. 
Firstly, the spread in radii of particles in the line of sight
is smaller at for lines of sight at larger distances from the center.
Secondly, the projection effects are less pronounced at larger radii.
Particles going at a given rotational speed have higher measured radial
velocities because of the inherent geometry. These points are
illustrated in Figure~\ref{fig:zvz2}.

\subsubsection{Model I: The Barred Galaxy Model}

We refer to  Model I in Table I as the barred galaxy
model.
For the case of this model, the situation
is much more complex. We have examined 3 cases which 
span the range of possible geometries. One with 
the bar at 45 degrees to the line of sight (referred to as bar45), 
one case with the bar aligned along the line of sight (bar0)
and thirdly a case where the bar is aligned perpendicular
to the line of sight (bar90). In each case we use the 
same galaxy model; we just change the viewing angle.
For each of these three cases we choose different 
inclination angles.

In the bar45 geometry viewed at an inclination of $60 \arcdeg$
we recover the true rotation curve with an accuracy 2~\kms. 
In the case of a 
bar aligned with the line of sight (bar0) one gets 
larger rotational velocities measured in the 
central kpc region of the galaxy than 
are truly present. This is  due to 
motions within the bar that are not rotational
but contribute to the perceived rotational 
velocity. Such a bar viewed lengthwise would 
not be classified as a bar but most likely as 
bulge and likely will be discarded. 
The velocity differences observed in this 
situation can be large; $\sim 10-15\kms$.
The third case when the bar is perpendicular
to the line of sight and clearly visible as a bar, results
in observed rotational velocities that are smaller than the 
true rotational velocities in the disk. 
The mismatch
between the long slit result and the rotational velocities
for a galaxy viewed at an inclination angle of $75^{\arcdeg}$ is 
of about 25\%
at a 1~kpc radius.
Inclination angle determinations are subject to rather 
large uncertainties as we mentioned above. If an inclination 
angle of 65$\arcdeg$ had been determined this galaxy the 
mismatch between the true and observed velocities
would be increased by several \kms. 
As we showed in Figure~\ref{fig:deblok}  it is galaxies 
with bars that tend to have the lower rotational velocities
in their centers.

The conclusion from the long slit analysis is that it is possible to 
underestimate the rotational motions in the central parts of dwarf 
galaxies by as much as 10~\kms\ using long slit data. This is a 30\%
error in the rotational velocity. The error  is simply 
due to the way the line of sight samples the velocities in a rotating
disk of finite thickness. Adding a bar into the picture can enhance
or decrease this effect depending on the orientation of the bar
relative to the observer. When the bar lies in the plane of the sky
the differences are increased. Bars are not always easy to detect.
However the inclination angle and position angle measured as a function 
of radius can give indications as to the presence of a bar. 

\subsection{The Tilted Ring Model: Results}
\subsubsection{Model IV  (no bar, no bulge)}

\begin{figure}[htb!]
\epsscale{0.99} \plotone{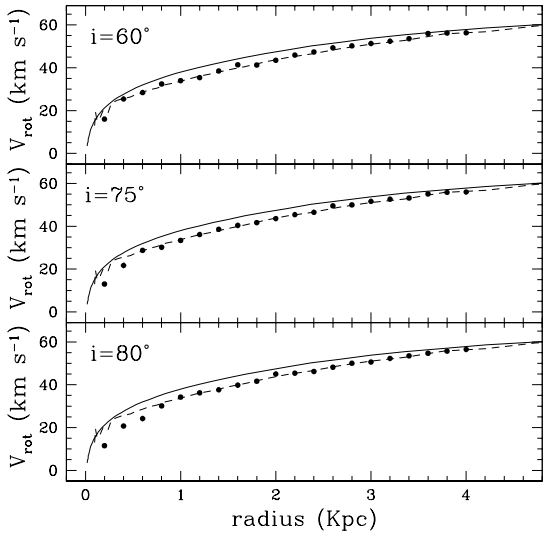}
\caption{Rotation curves for model IV, the
model with no bar and no bulge. The solid line shows the true circular
velocity.  The dashed line shows the rotational velocity. The points
show the ``observed'' velocities for the model viewed at inclinations
of 60\arcdeg, 75\arcdeg and 80\arcdeg. The effects of inclination are
very small for 60\arcdeg and for radii larger than 1~kpc. The mismatch
between the observed velocities and the true velocities at small radii
increases with increasing inclination.}
\label{fig:nobarinc}
\end{figure}

At low inclinations we find that a small error in the inclination measurement 
can affect the velocity results substantially. For example, we viewed 
model IV with at an inclination of 10\arcdeg~ but the ellipse fitting 
software estimated an inclination 14\arcdeg~resulting in underestimates 
of the rotational velocity of 10-15~\kms\ about 40\%. 
At inclinations of $\sim 60\arcdeg$
the measured velocities accurately match the rotational velocities over the 
measured range of 0.2 to 4~kpc. At an inclination of 75\arcdeg we find that
underestimates of the velocity of $\sim 5$\kms~ occur at radii of 
less than 0.4~kpc, a 20\% error in velocity. 
At an inclination of 80\arcdeg~ the velocity
underestimates occur at radii smaller than 0.7~kpc. At 0.2~kpc
the mismatch is about 8\kms, a 25\%
error in the velocity estimate. These results are shown in figure~\ref{fig:nobarinc}.

\subsubsection{Model I:Effect of a bar on the tilted ring results}

The tilted ring analysis described above is the one adopted
to calculate rotation curves using well resolved high signal 
to noise data. The method is applied to datasets where velocity 
information is available over the whole image plane, not just along 
a narrow slit. In practice observers have assumed a fixed or slowly 
varying position angle and inclination as a function of radius when 
implementing this method \citep{WdBW02}.
 
\begin{figure}[htb!]
\epsscale{0.99}
\plotone{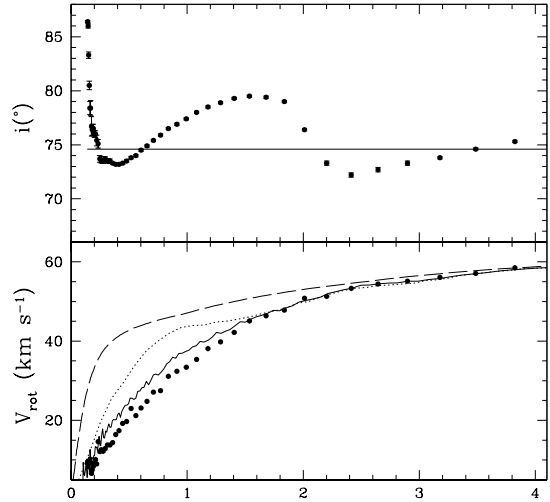}
\caption{Solid line in the bottom panel shows the average rotation 
velocity of particles in the disk as a function of radius. The dotted
line is the same as the solid line but particles belonging to the bar
have been excluded from the calculation. The dashed line  is the
circular velocity of particles in the disk. This is the velocity that 
particles would have if they moved in pure circular motion in the 
galaxy potential. At 0.5~kpc the measured velocity is 40\%
of the true circular velocity.}
\label{fig:fig8}
\end{figure}

To see the effect of the bar on rotation velocity, we selected 
disk particles in a ring from 1~kpc to 1.1~kpc and examined their 
rotation velocity as a function of azimuthal angle. The signature 
of the bar is clear.  Particles located close 
to the position angle of the bar have significantly smaller 
rotational velocities
on average. The bar is located  two quadrants of the 
galaxy. We compute the average velocity of all the 
particles in the ring  and obtained a value of 38.6\kms. When 
we exclude the quadrants containing the bar
we obtain a value of 43.5~\kms. We have  plotted the mean rotation 
velocities of particles in the disk, computed by excluding 
particles in the bar, this is the dotted line shown in 
Figure~\ref{fig:fig8}. We also show as a dashed line the circular 
velocity in Figure~\ref{fig:fig8}. The point is that there 
are a number of significant mismatches in the center of the 
galaxy. The tilted ring method does not recover the average
rotational velocities in the disk. Even if it did, this 
would not be sufficient, since the average is lowered by the 
presence of a bar. Even if one has a good measure of the rotational 
velocities in the disk, one needs an accurate method of calculating 
asymmetric drift to recover the circular velocity. 
The bottom line is that large mismatches occur between the observed
and true velocities in the central part of the disk when a bar is present. 
At 1~kpc from the center of the galaxy the observed velocity is 30\%
less than the true velocity. At 0.5~kpc from the center of the galaxy 
the observed velocity is 50\% of the true velocity.

It does not
take a very large error in velocity to produce a significant 
error in the power law slope of $\rho(r)$. Let us suppose 
we compute $\rho(r)$ at 1~kpc and 2~kpc from the center of the galaxy
and that we get the correct answer at 2~kpc but underestimate
the rotation velocity and 1~kpc. Let us say we think the velocity
is 31\kms\ when the true velocity is 35\kms. The percentage 
error is $\sim 10\%$. The rotational velocity enters as the 
square in the mass measurement so the resulting error 
in the density is $\sim 20\%$ such that a power 
law index -1 is measured to be -0.65.

\subsubsection{Model I: Recognizing a Bar}

\begin{figure}[htb!]
\epsscale{0.99}
\plottwo{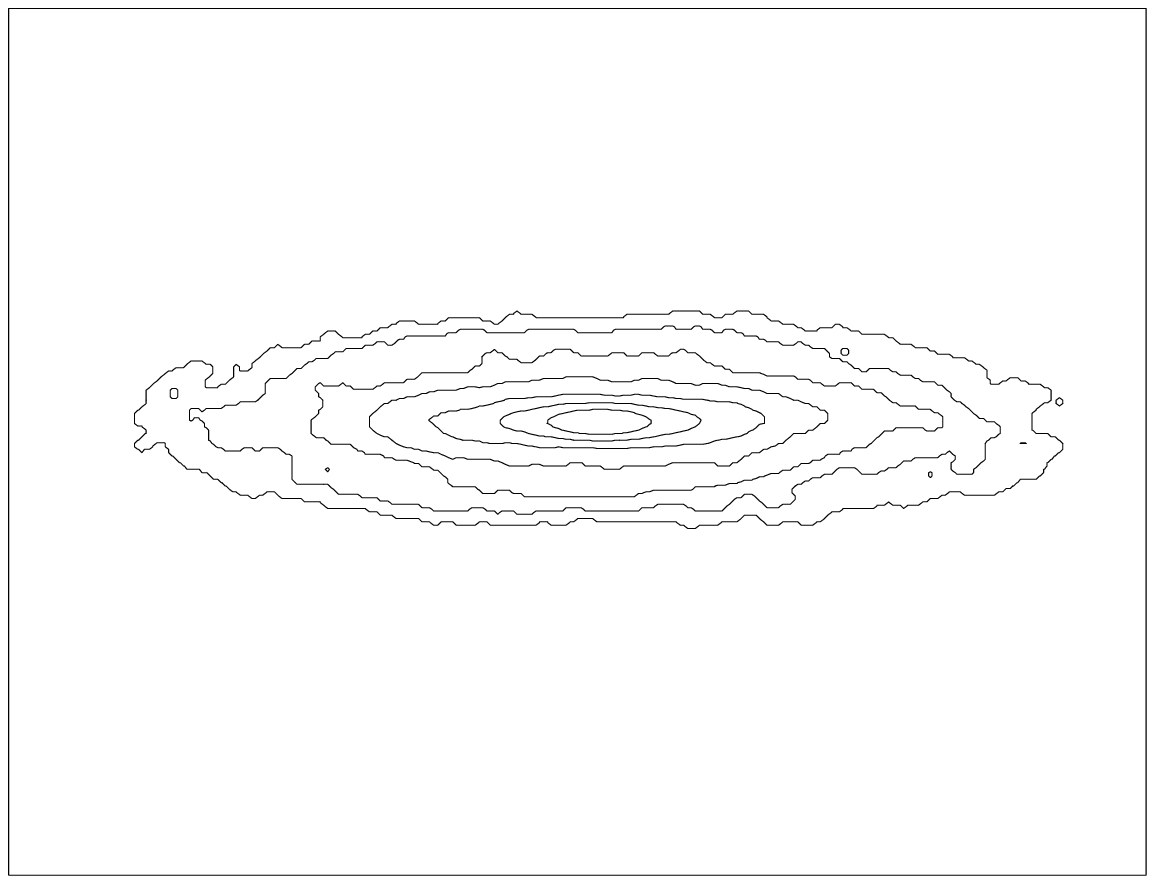}{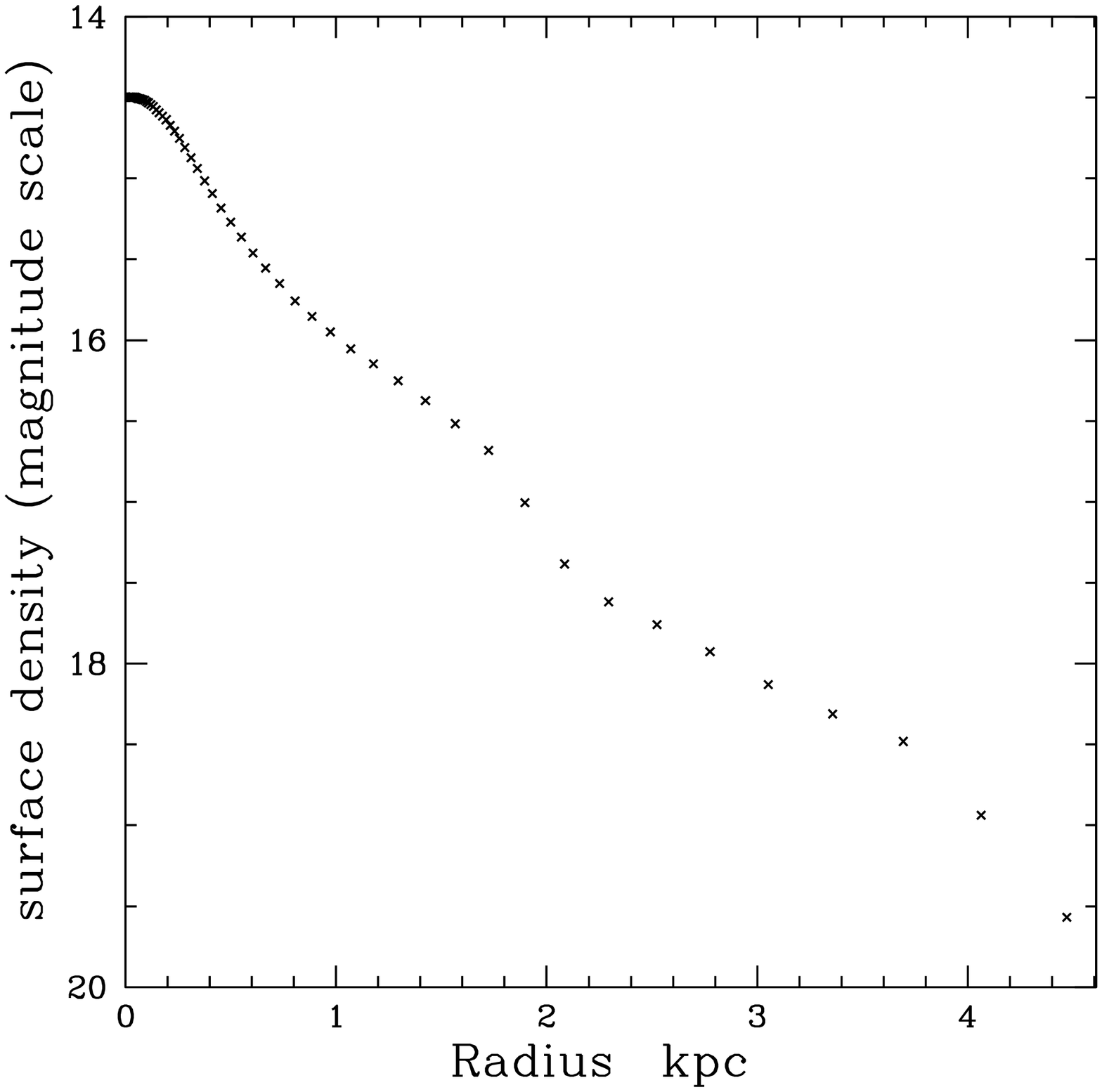}
\caption{Left hand side shows contour plot of the surface density of
particles in the 
barred galaxy viewed at an inclination of 75\arcdeg. The bar is
perpendicular to the line of sight.
Right hand side shows surface brightness versus semi-major axis length for the same model.
The surface brightness is measured on a magnitude scale.  Presence of the bar is hardly noticeable in
the contours of brightness. In the surface brightness the bars shows as a 0.5~mag feature at 2~kpc.}
\label{fig:contv}
\end{figure}

\begin{figure}[htb!]
\epsscale{0.99}
\plottwo{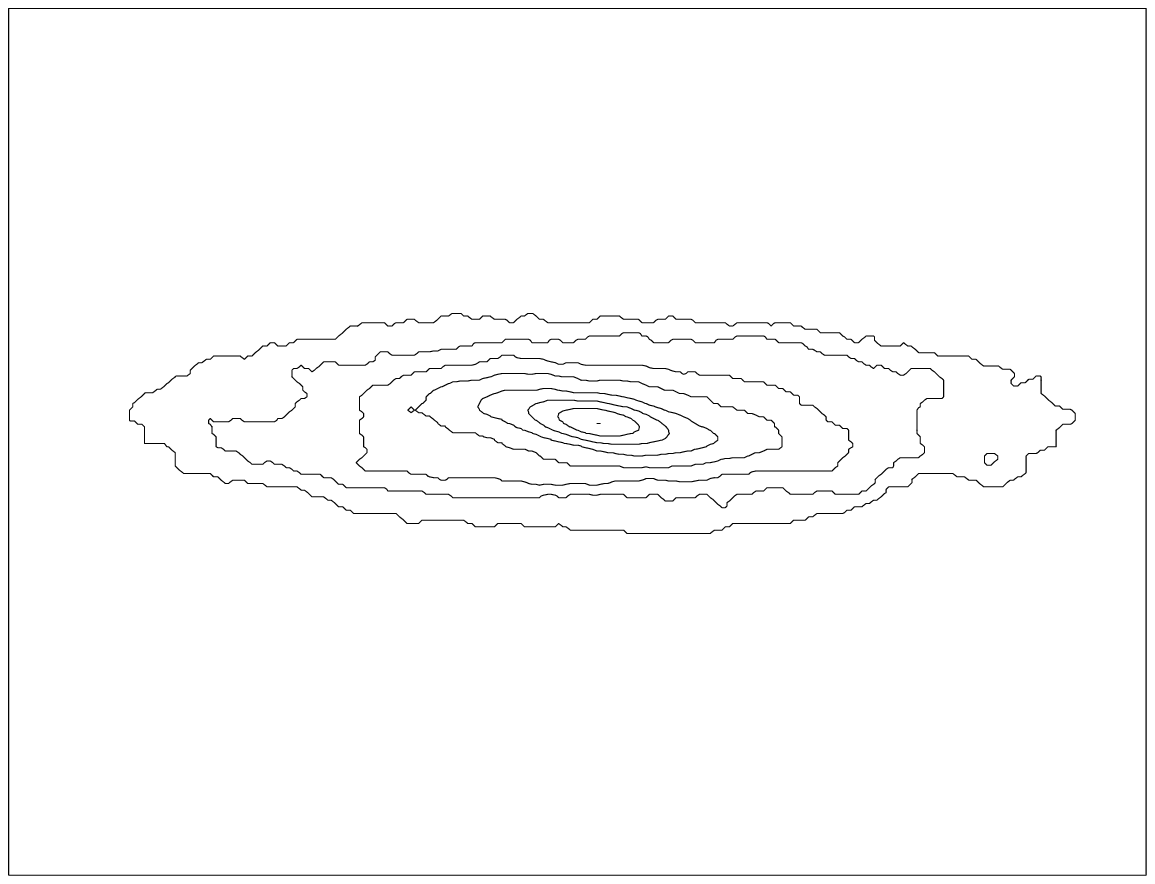}{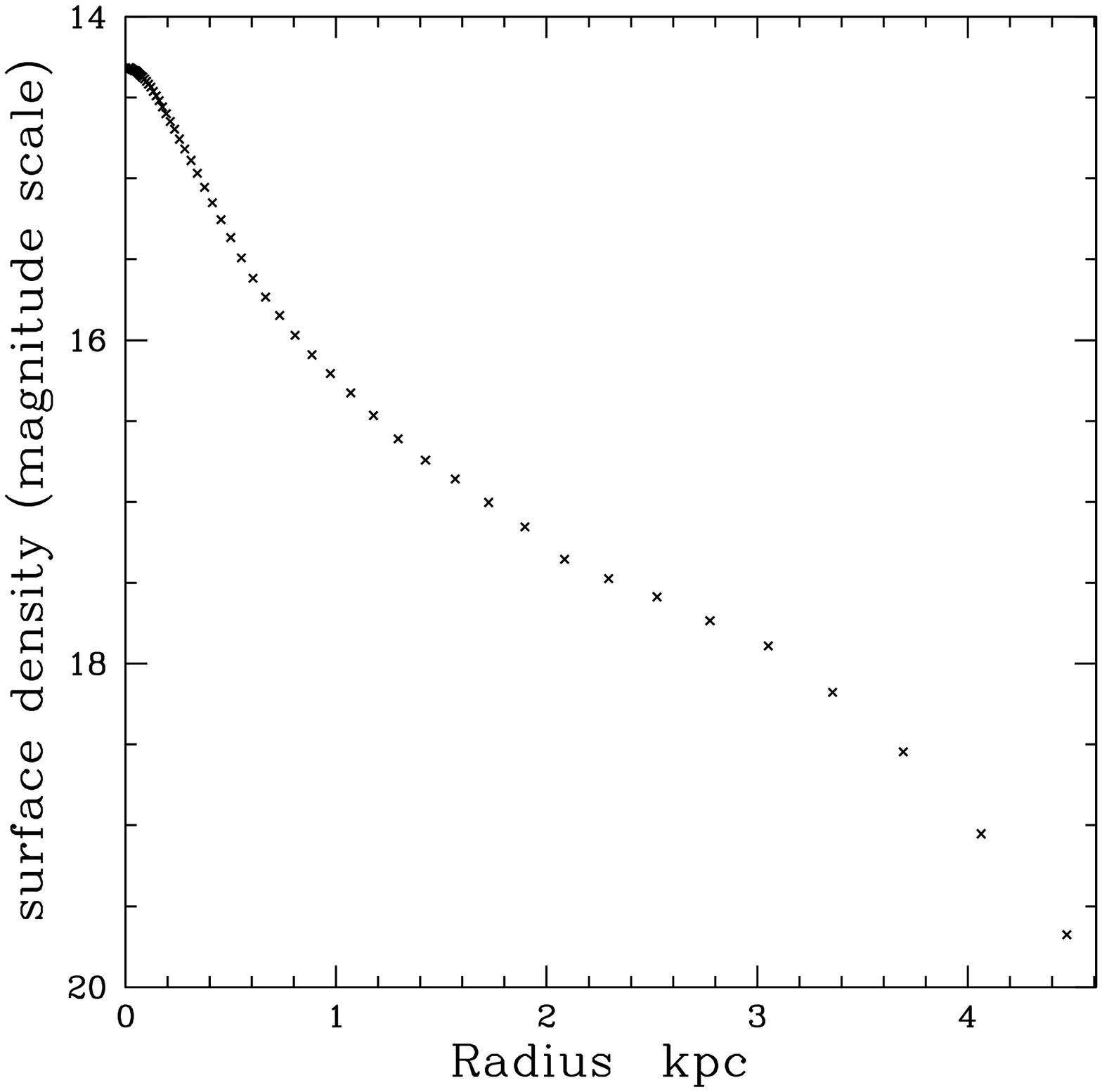}
\caption{Same as Figure~\ref{fig:contv}, except the azimuthal
angle of the bar is 45\arcdeg resulting in twisting isophotes. The bar
is not visible in the surface brightness profile. Presence of the bar is indicated only by  the change in the
position angle of isophots.}
\label{fig:cont45}
\end{figure}

CCD imaging studies reveal that $\sim$70\%
of spiral galaxies are barred  \citep{Esk02}. To avoid complications, galaxies should be 
selected for rotation curve studies from the remaining 30\%. 
However, our models suggest that bars may well be present in galaxies
where imaging does not reveal their presence.
We have demonstrated that bars in galactic disks give rise to errors
when inferring galaxy masses from rotation curves. One might argue 
that barred galaxies are easily recognizable and hence can be
easily excluded from galaxy samples. This is not the case. We show in figures \ref{fig:contv}
and \ref{fig:cont45} contour plots of the surface density of particles
for the barred galaxy model  viewed at an
inclination of 75\arcdeg. The surface density is measured as magnitudes
and the contours are equally spaced in magnitude to mimic observational
plots. It is certainly difficult to see the presence of a bar from
density contours alone. Bars do cause variations in the inclination
angle and position angles measured as a function of radius but warps 
and bulges can produce similar effects. We also show next to the
contour plots, the  ``luminosity profiles'' of our model disks.
These plots also reveal the presence of the bar in an indirect
way.

It is striking to what extent the signature of the bar can vary
in the surface density profile depending on the viewing angle.
The bar can have the appearance of a bulge, or produce a sudden
decrease in surface brightness, or have no marked effect.

\subsubsection{Model II:Evolution of a bar}

\begin{figure}[htb!]
\epsscale{0.99}
\plotone{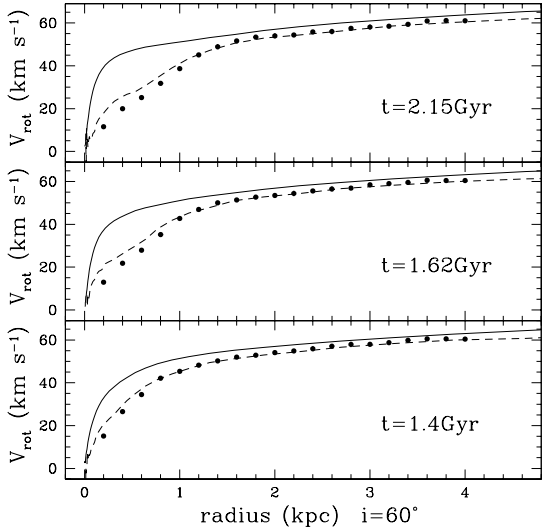}
\caption{Evolution with time of the mismatch between the circular
velocity and the observed rotation velocity.  The mismatch  gets
larger as bar grows.
Circular velocity (solid line), rotational velocity (dashed
line) and `observed velocity' (points) for the dwarf galaxy model
viewed during the formation and buckling of the bar. The model is
viewed at an inclination of $60\arcdeg$. The bar is oriented
perpendicular to the line of sight.}
\label{fig:3curve}
\end{figure}

\begin{figure}[htb!]
\epsscale{0.99}
\plotone{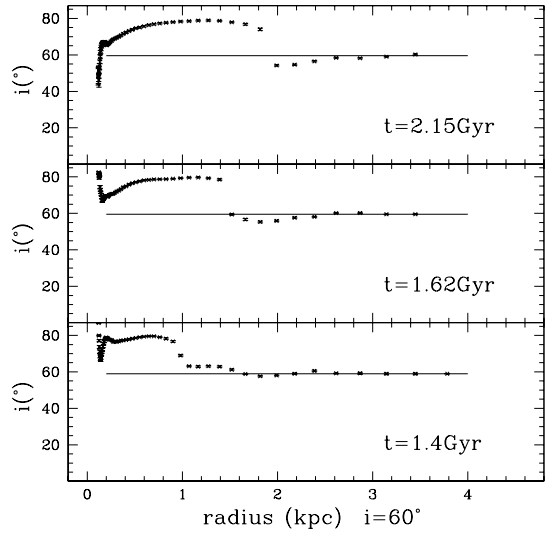}
\caption{Inclination angle as a function of radius for the dwarf
galaxy model viewed during the formation and buckling of the bar. The
model is viewed at an inclination of $60\arcdeg$. The bar is oriented
perpendicular to the line of sight.The solid line shows the true
inclination of the galaxy. The amplitude of the variation in the
inclination angle does not change with time, but the radius of the
transition to the true value increases with time and is a good measure
of the bar length.}
\label{fig:inc}
\end{figure}

We use model II to study effects of bar evolution on rotation curves.
We have examined outputs of the simulation at three
stages of the bar formation. The first snapshot was made 1.4~Gyr after
the start of the simulation when the bar is still weak. The second
snapshot was made 1.62~Gyr after the start of the simulation at which
point the bar is strong. The final snapshot was taken 2.15~Gyr after the 
start of the simulation when the bar has buckled. Buckling refers to the bending of the 
bar into a peanut shape forming a thickened double lobed structure
\citep{CS81}.
The rotation curve
results are shown in Figure \ref{fig:3curve}. As expected there is a large
increase in mismatch between the circular and observed velocities as
the strength of the bar increases. After the bar has buckled, the
observed velocity is less than 50\% of the circular velocity 0.5~kpc
from the center of the galaxy. Figure~\ref{fig:inc} shows the inclination 
based on the isophote analysis. One can clearly see the effect of the
bar extending to larger radii as the simulation evolves. Since we are
viewing the bar perpendicular to the line of sight there are no
isophote twists although these would be present if one viewed the bar
at some angle either than perpendicular or parallel to the line of sight.

\subsubsection{Model III: Effect of a bulge on the tilted ring results} 
 
\begin{figure}[htb!] 
\epsscale{0.99}
\plotone{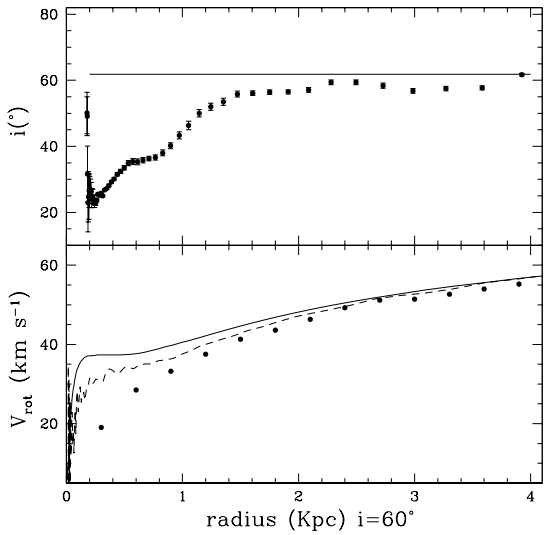}
\caption{Rotational velocity (bottom panel) and  inclination angle (top 
panel)  as a function of radius for model III (bulge)
viewed at an inclination angle of 60\arcdeg.
The solid line in the bottom panel shows the circular  velocity
of particles in the disk. The dashed line shows the rotational 
velocity and the points show the observed velocity.
The solid line in the upper panel 
is the mean inclination angle computed as the 
mean evaluated between 4 and 5~kpc.}
\label{fig:sbulge}
\end{figure}

Infrared studies of spiral galaxies and low surface brightness galaxies  
\citep{BSM03, GDIT02, BBH99} show that the majority of these systems
contain bulges. We have run a simulation  to explore what effect the
presence of a bulge has on the tilted ring results. Model III ``Bulge'' is 
designed to mimic the bulges detected by \citet{BSM03}: it has a very 
small bulge with half mass radius is 250~pc. Figure \ref{fig:sbulge} shows 
the results of the tilted ring analysis for model III.    
It is clear that bulges as well as bars can result in low rotational measurements in   
the central regions of galaxies where the circular velocity  is  in fact high.  
The differences are of order 10-20\kms\ in the central 0.5~kpc. The velocity error  
ranges from $\sim$20\% at 1~kpc to $\sim$ 50\% at 0.2~kpc. In the case of   a
barred model the greatest cause of the mismatch are the non-circular motions
produced by the bar. Due to the presence of the bar, a  large fraction of stars and gas parcels
have velocities orthogonal to the line of  sight. These velocities are undetected by 
the emission line observations. This effect strongly depends on the bar orientation  
and is  maximum when the bar is perpendicular  to the line of sight. The case of   
galaxies with small bulges is slightly  different. A bulge is a system substantially    
supported by random motions.  An observer  performed an average along the line  
of sight taking into account all particles belonging to either the disk or  the bulge.  
For disk particles the detected radial velocity is close to the projected circular velocity   
along the line of sight, a small correction should be applied because of the asymmetric  
drift. For bulge particles the detected radial velocity component is comparable to the velocity  
components orthogonal to the line of sight. Systematically bulge particles show  
smaller velocity along the line of sight compared with disk particles, for the same circular velocity.
Therefore the avenge or the observed rotation velocity is artificially biased toward 
lower values.

When models are viewed at low
inclinations, error in the inclinations can cause large systematic
errors in the rotation velocity estimates. For example the tilted 
ring algorithm produces an inclination value of $26.7\arcdeg$ for 
the model viewed at an inclination of $20\arcdeg$. The resulting 
error in $\sin i$ results in a $\sim$25\% error in velocity. 
A 25\% error in velocity translates as a 50\% error in the 
density in the outer parts of the galaxy.

The dynamical signatures of the 
bulge model extend to larger radii than its signature in the 
surface density profile plots. For an inclination of $60\arcdeg$, the
small bulge does not make a
significant contribution to the surface density profile beyond 400pc,
but rotation velocities are determined to be significantly lower due
to its presence beyond 1.2~kpc.  By comparison, the model 
with no bar and no bulge, model IV, when viewed at an inclination of 
$60\arcdeg$ shows good agreement between the rotation velocity and 
the observed velocity down to 200~pc.

\subsubsection{Model V: Effect of disk thickness on the tilted ring results} 
 
\begin{figure}[htb!]     
\epsscale{0.99}
\plotone{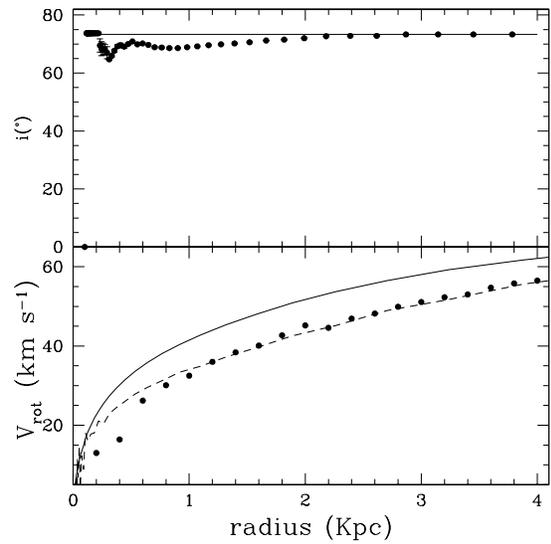}
\caption{Model V ``Disk'' observed at an inclination of 75\arcdeg.
The solid line shows the true circular
velocity.  The dashed line shows the rotational velocity. The points
show the ``observed'' velocities for the model.
The effects of inclination are
very small for 60\arcdeg and for radii larger than 1~kpc. The mismatch
between the observed velocities and the true velocities at small radii
increases with increasing inclination.}
\label{fig:thick}
\end{figure}

The motivation for Model V ``Disk'' is to gauge the effect of disk
thickness on the underestimation of rotational velocity in the central
Kpc of the models. Figure \ref{fig:thick} shows velocity and
inclination for the disk model viewed at an inclination of
$75\arcdeg$. The mismatch between the ``observed'' velocity and the
rotational velocity is 10~\kms~ at 0.4~Kpc and 6~\kms~ at 0.2~Kpc. The
thin disk model IV analysis reveals a smaller mismatch of 3~\kms~ at
0.4~Kpc and 6~\kms~ at 0.2~Kpc. As expected the mismatch increases with
disk thickness at the largest inclinations.

\subsubsection{Surface Density Profiles}

\begin{figure}[htb!]
\epsscale{0.99}
\plotone{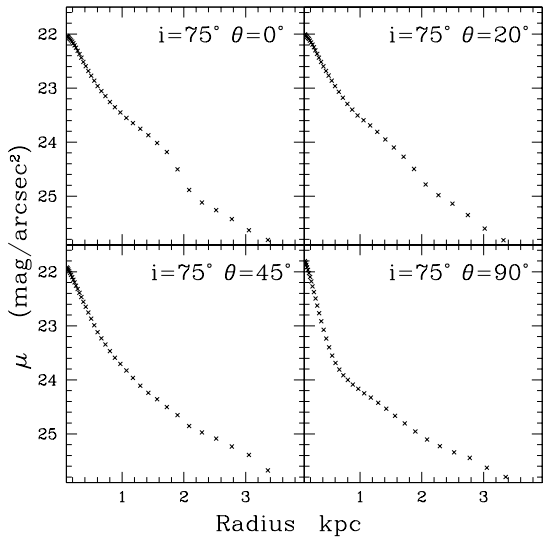}
\caption{The four panels show the barred galaxy model viewed at an
inclination of 75\arcdeg for varying angle that the 
bar makes with respect to the line of sight. In the upper left panel
the bar is viewed perpendicular to the line of sight. In the bottom
right panel the bar is viewed along the line of sight. The two other
panels are intermediate cases.}
\label{fig:panel}
\end{figure}

Using the Ellipse task in the package STSDAS we can produce surface
density profiles for our models.  This is the same package
used by \citet{BSM03} among others to produce surface density profiles
of galaxies. We are thus processing our $N-$body ``data'' in as similar a
manner as possible to the observers.  The surface density profile 
for the barred galaxy model is show in the four panels of
Figure \ref{fig:panel}. We are viewing the galaxy at an inclination of
75\arcdeg and showing how the surface density profile is affected by
varying the angle the that bar makes with the line of sight. The bottom 
right panel showing the profile that results in viewing the bar along
the line of sight, illustrates the point that in this geometry a bar
can be mistaken for a bulge. When the bar is viewed perpendicular to
the line of sight we see a sudden dip in the surface brightness at
about 2~kpc. This dip is not so prominent when the bar is only 
slightly rotated from the perpendicular (upper right panel).
It is clear that a number of the features found in these surface
brightness plots qualitatively resemble features found in the 
data presented by \citet{BSM03}.

\begin{figure}[htb!]
\epsscale{0.99}
\plotone{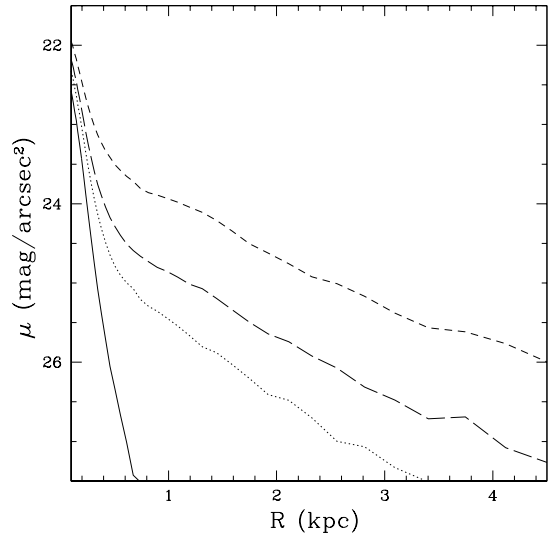}
\caption{Surface brightness profiles for model III (bulge model)
viewed at an inclination of 20\arcdeg (dotted line) 60\arcdeg (long
dashed line) 80\arcdeg (short dashed line).  The solid line shows the
surface brightness profile of the bulge on its own. The bulge
contribution to the observed profile gets smaller as the inclination
increases. Surprisingly, such a small bulge affects the rotation curve
and the measured inclination angle up to 1~kpc as indicated in
figure~\ref{fig:sbulge}.}
\label{fig:nopanel}
\end{figure}

Figure \ref{fig:nopanel} shows the same analysis for model III (bulge model).
By varying the inclination angle we vary the contribution of
the disk to the surface brightness. The slope of the disk contribution
also varies depending on the viewing angle. The bulge signature in the 
surface brightness plot does not extend beyond $\sim 0.6$~kpc whereas the 
rotation velocity is underestimated out to $\sim 1.2$~kpc.

\begin{figure}[htb!]
\epsscale{0.99}
\plotone{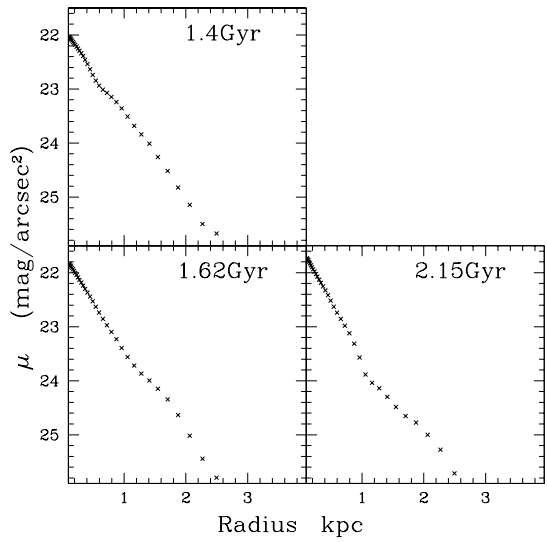}
\caption{Dwarf galaxy model viewed at varying times during its
evolution. 1.4~Gyr corresponds to the beginning of bar formation. By
1.62~Gyr a strong bar has formed, at 2.15~Gyr the bar has buckled as
described in the text. The bar is viewed at 45\arcdeg to the line of
sight in each instance}
\label{fig:paneldwarf}
\end{figure}

Figure \ref{fig:paneldwarf} shows the effect of viewing the model at constant
inclination during its evolution. One can see the growth of the bar as
a shift in the location of the point at which the profile slop changes.
This point occurs at roughly 0.5~kpc in the upper panel, shifts to
1.5~kpc in the lower left panel and back inward to about 1~kpc as the
bar buckles.

Figures \ref{fig:panel} to \ref{fig:paneldwarf} illustrate the variety
of profiles that can be produced by adding bulges and bars to rotating
disks. In few cases, based on surface density profiles alone,  
is it immediately clear that a bar or bulge must be 
present. Bars can appear as bulges or disappear almost entirely
depending on the viewing angle. Bulges can affect only the central
200~pc or extend as far as 1~kpc depending on the intrinsic bulge size
and viewing angle. The bar signature also changes as the instability
evolves in the disk. We have shown earlier that bars and bulges
significantly affect the interpretation of rotation curves. The figures 
discussed in this section show that is is difficult to unambiguously
identify these features based on surface density profiles alone.
This section is intended as a cautionary note to researchers who ignore
these features in the analysis of galaxy rotation curves.

\subsection{Density Profiles From Rotation Curves}  
As explained in section 5.2, one can infer the density distribution in a galaxy   
from its rotation curve using eq. 12.  Under the assumption that gas and stellar  
mass contribution is negligible,  it is possible to recover the dark matter halo  
density profile. LSB and Dwarf  galaxies are considered to closely satisfy the    
assumption of being completely dark matter dominated and it is the application of 
this is the method to LSB and Dwarf galaxies which gives rise to the inference of  flat  
cores in the central density  distributions of galaxies. With our models we can 
check the accuracy  of  this method.

\begin{figure}[htb!]
\epsscale{0.99}
\plotone{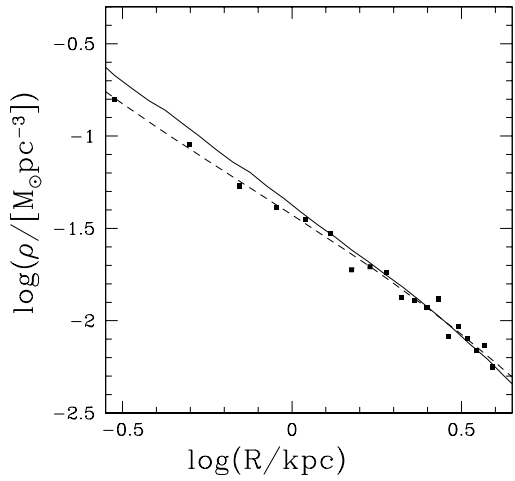}
\caption{Density versus radius for model IV (no bar, no bulge)
viewed at an inclination angle of 80\arcdeg.
The solid line is the actual density of the model.  
The points show the observed density and the dashed line 
is a fit to the observed density.}
\label{fig:dnb}
\end{figure}

For  model IV (with no bar and no bulge) the ellipse fitting method does a
reasonable job recovering the true density. The density versus 
radius is shown for the model with no bar viewed at an inclination 
of 80\arcdeg\ in Figure \ref{fig:dnb}. At radii of less than 1~kpc
the density is underestimated by about 30\%. 
The underestimates vary from 10\%\ to 50\% 
depending on the inclination angle.

\begin{figure}[htb!]
\epsscale{0.99}
\plotone{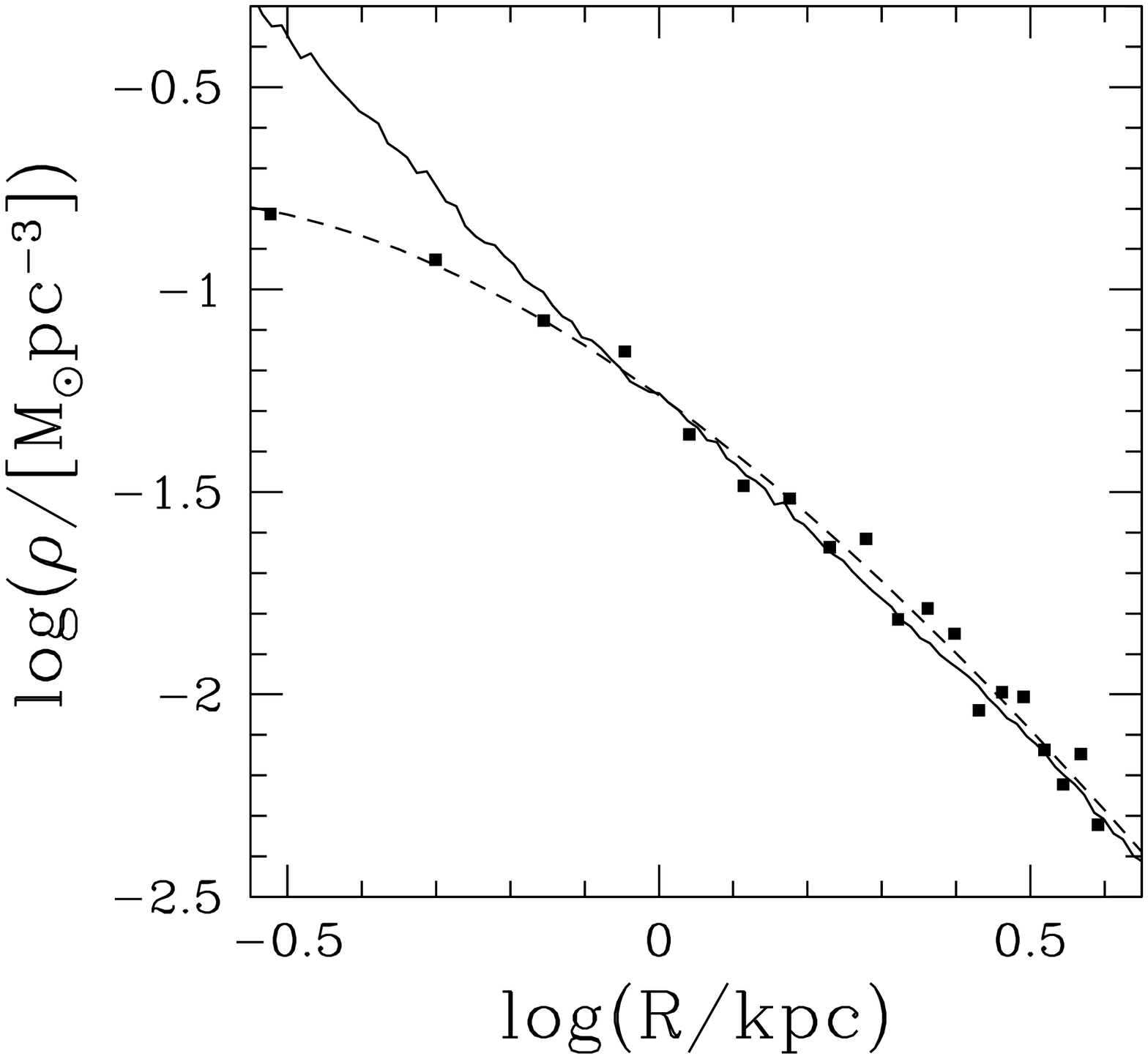}
\caption{Density versus radius for model I (barred model)  
viewed at an inclination angle of 75\arcdeg.
The solid line is the actual density of the model.
The points show the observed density and the dashed line 
is a fit to the observed density. The orientation of the bar
is 45\arcdeg to the line of sight.}
\label{fig:rb75}
\end{figure}

\begin{figure}[htb!]
\epsscale{0.99}
\plotone{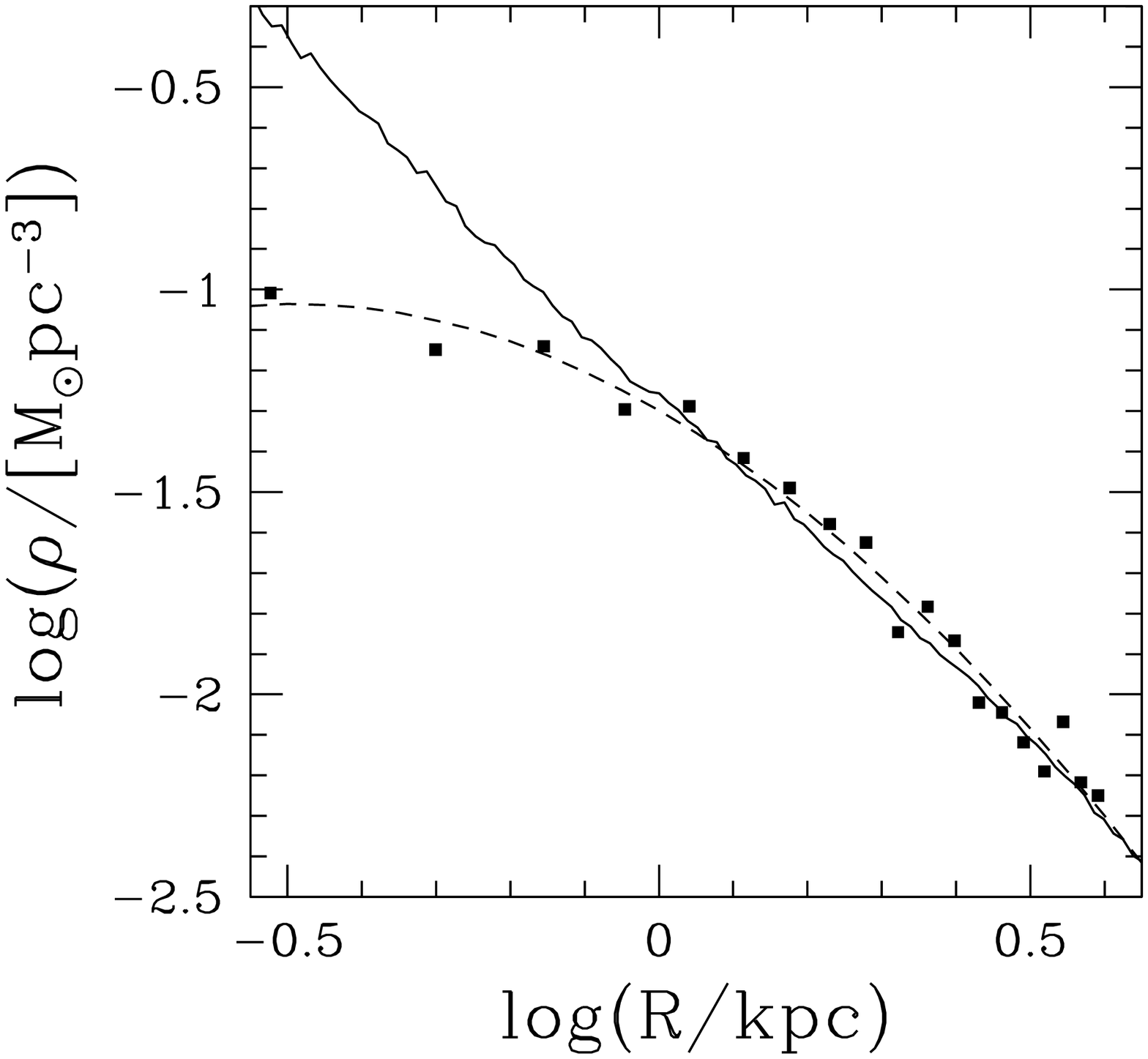}
\caption{Density versus radius for model I (barred model)
viewed at an inclination angle of 75\arcdeg.
The solid line is the actual density of the model.
The points show the observed density and the dashed line 
is a fit to the observed density. The bar is perpendicular
to the line of sight.}
\label{fig:rb75v}
\end{figure}

\begin{figure}[htb!]
\epsscale{0.99}
\plotone{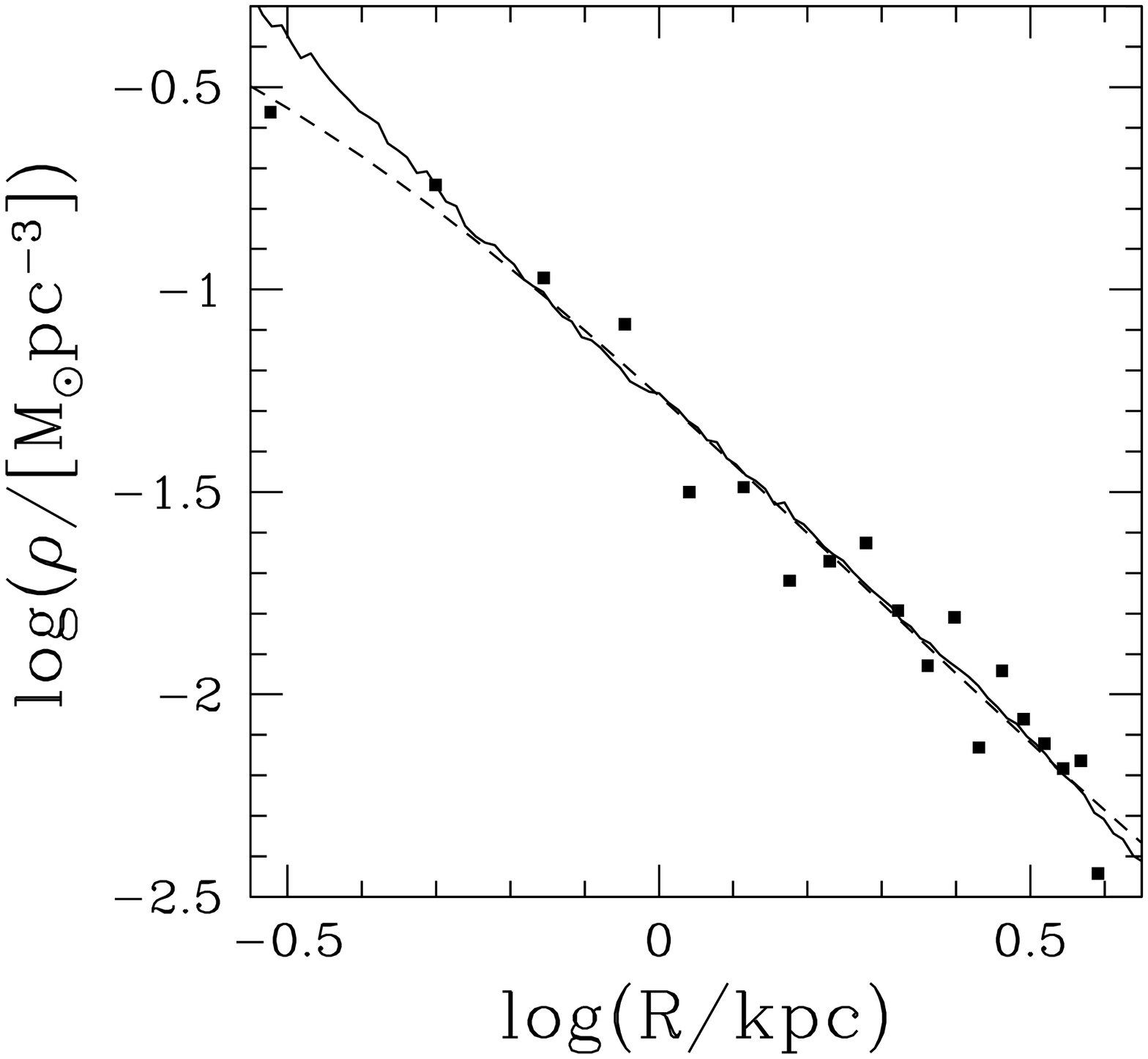}
\caption{Density versus radius for the barred model  
viewed at an inclination angle of 75\arcdeg.
The solid line is the actual density of the model.
The points show the observed density and the dashed line 
is a fit to the observed density.
The bar is viewed along the line of sight.}
\label{fig:rb75h}
\end{figure}

These effects are much more pronounced for the barred galaxy model.
In figure \ref{fig:rb75} we show the density for the barred model 
viewed at an inclination of 75\arcdeg.  
Figure \ref{fig:rb75v} and Figure \ref{fig:rb75h} show the same analysis 
with the bar in the plane of the sky and perpendicular to the plane 
of the sky. The effects are most pronounced on the latter case.
When the bar is perpendicular to the line of sight it is easy to see 
how an observer would infer the presence of a constant density core 
in the center of the galaxy. The orientation of the bar with respect to  
the line of sight has a strong effect on the inferred density profile. 

\begin{figure}[htb!]
\epsscale{0.99}
\plotone{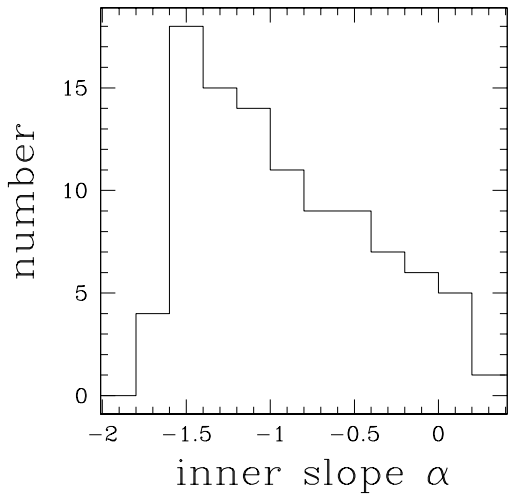}
\caption{Histogram of the slopes obtained from 
model I (bar) viewed at 
inclination angles of 80 75 and 60 \arcdeg. 
The slopes where measured at positions evenly spaced between
0.2 and 1~kpc. The true slope of the model is -2.}
\label{fig:hist}
\end{figure}

We have computed the the slope of the $\rho(r)$ curve at locations  
smaller than 1~kpc for the three barred models. We show these in the  
histogram in Figure \ref{fig:hist}. The key points is that for barred  
galaxy models viewed at an inclination greater than or equal to  
60\arcdeg we {\it always} measure a slope that is shallower  
than the true slope. For a true slope of -2 we can obtain slopes  
as shallow as 0. Note that the NFW profile has a slope of -1, 
in the light of our findings it is not surprising that slopes  
of shallower than -1 are being found near the centers of  
dwarf galaxies.

\section{Conclusion}

The rotation curves of dwarf and LSB galaxies arguably present the most
challenging problem for current cosmological models. The key to this problem
lies in reliably estimating the accuracy with which 
one can recover the mass distribution and, thus, the dark matter
density using observed rotation curves. This is the main question
which we address in this paper by studying realistic numerical models.
The models mimic real galaxies in many respects. For example, they
have thin ($\sim 100~$pc) and cold ($\sim 10~\kms$) exponential
disks. The latter condition is very important because many effects are related
to the magnitude of random motions.

The main lesson to be learned from our analysis is that there are
many systematic biases.  Each bias can be very small (only few km/s),
yet they all add up to produce significant effects.  We find that the tilted ring analysis of
spectroscopic data results in large 30\% to 50\% underestimates of the
rotation velocity in the central 1~kpc region of the galaxy
models. These  underestimates in velocity in turn result in the inference
of slopes in the central density distributions of our models that are
shallower than the true slopes.

A quantitative understanding of these biases is important 
if one is to accurately interpret
rotation curves. If one ignores these biases, the analysis 
of rotation curves can produce misleading results.
The extent to which this has occurred for real galaxies can only
be resolved by detailed analyses of individual objects. One cannot 
be sure that the analysis of real observations data was done properly when the
same analysis produces false results when applied to realistic models.

One of the difficulties in interpreting observational data lies in 
the fact that galaxies are treated in an overly simplified manner (e.g., cold gas
inside thin exponential disks). Observational results (including those 
presented in this paper) suggest that this is far too naive a picture. In
the central region, the stellar component often rotates as fast as the
emission-line gas - in contradiction with the prediction of a naive picture of 
a cold gas embedded in a hot stellar component.  The presence 
small (200-500~pc) bulges and bars, which can be easily overlooked,
further complicates the interpretation of the data.

\acknowledgments 

We acknowledge financial support of NSF and NASA grants to NMSU. Our
simulations were done at the National Center for Supercomputing
Applications (NCSA), at the National Energy Research Center (NERSC),
and at the Direcci\'on General de Servicios de C\'omputo Acad\'emico,
UNAM, Mexico, whose support we also acknowledge.  GR is thankful to
NMSU for hospitality.  We are grateful to R.Walterbos, J.Dalcanton ,
J. Gallagher, and E. Athanassoula for helpful discussions.


\clearpage


\clearpage 

\clearpage

\begin{deluxetable}{lcclllll}
\tablewidth{459.8pt}
\tablenum{1}
\tablecolumns{8}
\tablecaption{Parameters of galaxies with measured rotation curves}
\tablehead{\colhead{Galaxy} & \colhead{M$_B$}  & \colhead{$V_{LG}$} & \colhead{Type}
& \colhead{Incl.} & \colhead{$R_{\rm disk}/R_{\rm bulge}$} & \colhead{Scale} & \colhead{Comments} \\
 & & $km/s$ & & deg & kpc & kpc$/^{``}$\\}
\startdata
IC 267 & -19.91 & 3577  & SBb   &45.0  & 2.0/ 0.4 & 0.248 & Very long 10~kpc bar. Compact\\
       &&&&&&&  bulge.\\
IC 1029& -21.00  & 2520  & Sb   &76.0 & 5.0/ 0.4  & 0.175 & Compact bulge. Gas rotation is\\ 
       &&&&&&& measured using SII emission line.\\ 
NGC 5389& -19.17  & 1996  & SABa   &78.5  &3.0/ 0.45  & 0.14 & Strong ring at 6~kpc. \\
NGC 5746& -20.51  & 1676  & SABb   & 80.5 &2.5/ 1.2 & 0.12 & Large peanut-shape bulge/bar \\
NGC 5719& -18.79  & 1676  & SABab   & 70.0  &2.2/ 1.1 & 0.116 & Barred galaxy with a large bulge. \\
NGC 5793& -19.11   & 3387  & Sb    &70.5 &4.2/ 1.1 & 0.234 & \\
NGC 6368& -19.98   & 2904  & Sb   &75.5 &6.0/ 0.7 & 0.201 &  Gas rotation is measured using \\
       &&&&&&& SII emission line.\\
NGC 7537& -9.42   & 2888  & Sbc  &70.0 &5.5/ 1.4  & 0.200 & \\ 
\enddata
\end{deluxetable}

\clearpage

\begin{deluxetable}{llllll}
\tablewidth{459.8pt}
\tablenum{2}
\tablecolumns{6}
\tablecaption{Parameters of the Models}
\tablehead{\colhead{Parameter \ Model} & \colhead{Model I}  & \colhead{Model II} & \colhead{Model III}
& \colhead{Model IV} & \colhead{Model V}\\
 & ``Bar'' & ``Dwarf'' & ``Bulge'' & ``Thin Disk''& ``Disk''}
\startdata
Halo mass M$_{\rm halo}, \msun$  \phm{0} & 4.3$\times 10^{10}$ &6.0$\times 10^{10}$     &6.8$\times 10^{10}$  &6.0$\times 10^{10}$ &6.0$\times 10^{10}$\\
Disk mass M$_{\rm disk}, \msun$  \phm{0}  & 9.5$\times 10^8$ & 5.7$\times 10^8$  &6.8$\times 10^8$  &5.7$\times 10^8$ &5.7$\times 10^8$ \\
Bulge mass M$_{\rm bulge},\msun$  \phm{0} &  --     &  --   & 8.2 $ \times 10^{7}$   & -- & -- \\
Initial disk scale length $R_d$, kpc \phm{0} & 1.16   & 0.7   & 1.75   & 2.14 & 1.75\\
Final disk  scale length $R_d$, kpc  \phm{0} & 1.50   & 0.7   & 1.75   & 2.14 & 1.75\\
Initial disk scale height  $z_d$, pc \phm{0} & 90   & 108  & 55   & 150 & 300 \\
Final disk scale height $z_d$, pc \phm{0} & 135   & 120   & 70    & 150 & 300\\
Disk stability parameter $Q$ \phm{0} & 1.2 & 1.2  & 1.5  & 3.0 & 3.0\\
Bulge half-mass radius, pc \phm{0}  &   --    &  --  & 250 & --  & -- \\
Number of halo particles  \phm{0}    &3.3$\times 10^6$  & 3.45$\times 10^6$ &2.2$\times 10^6$  &3.45$\times 10^6$&3.45$\times 10^6$\\
Number of disk particles \phm{0}      &  2 $\times10^5$ & 1.6$\times10^5$ &  8$\times10^4$  &1.6$\times 10^5$&1.6$\times 10^5$\\
Number of bulge particles  \phm{0}  &  --     &  --    & 9.7 $\times 10^{3}$ & -- & -- \\
Halo concentration C \phm{0}  & 11.5                                         & 12.0                    & 11.0 & 12.0& 12.0\\
Maximum Force Resolution, pc \phm{0} & 11  & 16 & 19  & 16 & 16\\
Duration of evolution, Gyrs & 4.2 & 2.2 & 1.4 & 1.6 & 1.2\\
Time-step, yrs & $4\times 10^4$ & $1.4\times 10^4$ &$9.1\times 10^4$  & $1.4\times 10^4$ & $1.4\times 10^4$ \\
\enddata
\end{deluxetable}

\clearpage

\end{document}